\documentclass[11pt,a4paper]{article}
\usepackage{axodraw}
\usepackage[utf8]{inputenc}
\usepackage{epsfig}   
\usepackage{amssymb}
\usepackage{latexsym}
\usepackage{amsmath}
\usepackage{graphics,subfigure} 
\usepackage{multirow}
\usepackage{color}   
\usepackage{hyperref}
\usepackage[left=2.5cm,right=2.5cm,top=2.5cm,bottom=2.5cm]{geometry}

\newcommand{\lsp} {\tilde{\chi}_1^0}
\newcommand{\mlsp} {m_{\tilde{\chi}_1^0}}

\def\chia{\tilde{\chi}_1^0}

\def\chiapm{\tilde{\chi}_1^{\pm}}

\def\br{\mathrm{BR}}

\newcommand{\tone} {{\tilde t}_1}

\newcommand{\mtone} {m_{{\tilde t}_1}}

\newcommand{\mbone} {m_{{\tilde b}_1}}

\newcommand{\msql} {m_{{\tilde q}_L}}
\newcommand{\mlepl} {m_{{\tilde \ell}_L}}

\newcommand{\mgl} {m_{\tilde g}}

\newcommand{\mnulepl} {{\tilde m}_{\nu_L}}

\newcommand{\mtauone} {m_{{\tilde \tau}_1}}

\newcommand{\tanb} {\tan\beta}

\newcommand{\hone} {h_{1}}

\newcommand{\htwo} {h_{2}}

\newcommand{\hpm} {h^{\pm}}

\newcommand{\ra} {\rightarrow}

\newcommand{\lsim}{\raisebox{-0.13cm}{~\shortstack{$<$ \\[-0.07cm] $\sim$}}~}
\newcommand{\gsim}{\raisebox{-0.13cm}{~\shortstack{$>$ \\[-0.07cm] $\sim$}}~}
\newcommand{\bea}{\begin{eqnarray}}
\newcommand{\eea}{\end{eqnarray}}
\newcommand{\beq} {\begin{equation}}
\newcommand{\eeq} {\end{equation}}

\newcommand{\nn}{\nonumber}
\newcommand{\order}{\cal O}

\title{Production of a light top-squark pair in association with a \\ light non-standard Higgs boson within the NMSSM at the 13 TeV LHC  and a 33 TeV proton collider}

\author{Siba Prasad Das\footnote{Email: sp.das@uniandes.edu.co}, Jorge Fraga\footnote{Email: jf.fraga@uniandes.edu.co} and Carlos Ávila\footnote{Email: cavila@uniandes.edu.co} \vspace{3mm} \\
 \textit{Department of Physics, Faculty of Science,} \\
\textit{Universidad de los Andes, Apartado Aéreo 4976-12340, Carrera 1 18A-10,} \\
\textit{Bogotá - Colombia.}}

\date{\today \\}

\begin{document}

\maketitle

\begin{abstract}
    We study the potential of the LHC accelerator, and a future 33 TeV proton collider, to observe the production of a light top squark pair in association with the lightest Higgs boson ($\tone\tone\hone$), as predicted by the Next-to-Minimal Supersymmetric Standard Model (NMSSM). We scan randomly about ten million points of the NMSSM parameter space, allowing all possible decays of the lightest top squark and lightest Higgs boson, with no assumptions about their decay rates, except for known physical constraints such as perturbative bounds, Dark matter relic density consistent with recent Planck experiment measurements, Higgs mass bounds on the next to lightest Higgs boson, $\htwo$, assuming it is consistent with LHC measurements for the Standard Model Higgs boson, LEP bounds for the chargino mass and Z invisible width, experimental bounds on B meson rare decays and some LHC experimental bounds on SUSY particle spectra different to the particles involved  in our study. We find that for low mass top-squark, the dominating decay mode is  $\tone \rightarrow b \chiapm$ with $\chiapm  \rightarrow W \chia$. We 
    use three bench mark points with the highest cross sections, which naturally fall within the compressed spectra of the top squark, and make a phenomenological analysis to determine the optimal event selection that maximizes the signal significance over backgrounds. We focus on the leptonic decays of both $W$'s  and the decay of lightest Higgs boson into b-quarks ($\hone\rightarrow b \bar b$). Our results show that the high luminosity LHC will have limitations to observe the studied signal and only a proton collider with higher energy will be able to observe the SUSY scenario studied with more than three standard deviations over background.   
\end{abstract}

\pagestyle{plain}

\section{Introduction}
\label{sec:intro}

It is expected since long that the mechanism that triggers the electroweak symmetry breaking (EWSB) and generates the fundamental 
particle masses will have at least two parts \cite{Ellis:2017upx}. The first one is the search and the observation of a spin-zero 
Higgs particle that will confirm the scenario of the minimal Standard Model (SM), which has one Higgs isospin doublet, of 
Glashow-Weinberg-Salam and most of its extensions, that is, a spontaneous symmetry breaking by a scalar field that 
develops a non-zero vacuum expectation value (VEV). This part has recently been discovered by the ATLAS and CMS experiments at 
the Large Hadron Collider (LHC) with the observation of a new boson with a mass of around 125 GeV \cite{higgs125}. This is 
apparently consistent with the symmetry breaking mechanism in the SM and opens up the second part, which is mandatory to 
establish the exact nature of the symmetry breaking mechanism and, eventually, identify the possible effects of 
Supersymmetry (SUSY). The available SUSY models at our disposal are many: minimal Supergravity (mSUGRA), 
Minimal Supersymmetric Standard Model (MSSM) \cite{Kane:1993td}, 
Next-to-MSSM  (NMSSM)\cite{nmssm,Drees:1988fc,Franke:1995tc,Maniatis:2009re,review}, etc.
In our analysis we consider NMSSM model which solved the $\mu$-problem of the MSSM. It is to be noted that the earlier development for 
the origin of the $\mu$-term in Supertpoetial studied from Supergravity approach \cite{fayet} and from the non-renormalization operators \cite{gfam}.

The appealing of SUSY over the SM leads to an enlarged particle spectrum. As none of the SUSY  
particles has been discovered by any high energy collider experiment yet,  SUSY must be broken.
Various scenarios of the breaking mechanism have been developed in the past few decades. 
The  top quark, the heaviest particle among the SM quark sector, having a large Yukawa coupling, has a superpartner  called
the top-squark, which is expected to be the lightest among all the sparticles in the enlarged squark sector, except for the gaugino sectors, like neutralinos and charginos. 
The light top-squark is theoretically 
favored from the stabilization of the Higgs potential, as well as from the longitudinal scattering 
of the $W$-boson \cite{Ishiwata:2017rhf}.
Since it contains color charge, it is expected to be produced in abundance in the present and upcoming hadron colliders. 
The current operating LHC has already excluded the lighter top-squark for masses of around 800 GeV \cite{cms2017, atlas2017}.
However, these limits are under the assumption that only one decay mode of the top-squark is dominating. From the SUSY model perspective, this assumption is realized only in a very special range of the model space. Using this 
kind of non-observations, leads to  bias  exclusion limits of the model spaces  
\cite{deJong:2012zt,Ellwanger:2014dfa,Bagnaschi:2017tru}.
In our study, we are relaxing this criterion of a single dominant decay mode, and rather take
into consideration the branching ratios of all the allowed decay modes in the allowed model spaces. 

{\it Generic top-squark phenomenology:}
It seems to us that probably \cite{Djouadi:1997xx} is the most earliest work on the associated top-squark pair 
production with Higgs boson phenomenology (within MSSM) much before the discovery of the SM-Higgs boson. 
The author considered the mass of the top-squark is not that much larger than the top-quark (somewhat similar 
to the compressed scenario),  which later from the LHC searches founds the most difficult scenario to 
disentangle the lighter top-squark signal. In our scenario, we considered this kind of mass ranges of the 
lighter top-squark however the Higgs boson is replaced by the non-Standard type. The couplings of the 
top-squark and hence the production cross-sections within the NMSSM  would be very different from the earlier 
works and is interesting to explore with the  present generation and future collider 
experiments. In another work the authors extended the analysis for energy upgraded 
hadron colliders and $e^+ e^-$ colliders \cite{Djouadi:1999dg}. 

The lighter top-squark phenomenology from the SUSY dark matter perspective has had much interest since decades 
\cite{Boehm:1999bj,Baer:2017pba}. The top-squark phenomenology at the LHC has been studied recently from radiatively-driven 
naturalness and light Higgsinos scenario  \cite{Baer:2016bwh,Baer:2017yqq}, and from the bottom-squark pair production \cite{Han:2013kga,An:2016nlb}. The cosmological perspective, with direct and indirect detection of dark matter with top-squark co-annihilation,
has been studied recently in \cite{Pierce:2017suq}. 
In the compressed mass scenario, generic SUSY phenomenology has been studied by \cite{Florez:2016lwi}, \cite{An:2015uwa}, \cite{Kang:2017rfw}, \cite{Cheng:2017dxe}. In this latter case, the decay patterns of the lighter top-squarks change dramatically with minimal mass variations. Thus, the branching ratios are very different and the event rates in any particular signal channel change considerably.
Finding the lighter top-squark \cite{Yamanaka:2014pua} together with a Higgs boson has been studied for  many years \cite{Heath:2009gi}. Non-Standard Model Higgs searches studies have been made recently in the context of a Large Hadron-electron collider \cite{Das:2015kea}, \cite{Das:2016eob}. On the other hand, studies on the non-standard Higgs boson at the LHC have been made  in \cite{Ellwanger:2017skc}, from the decay of heavy Higgs boson to lighter non-standard Higgs bosons. In \cite{Baum:2017gbj} the author studied the mono-Higgs 
signatures from the Higgs boson to Higgs bosons decays within the NMSSM model. They found that heavy Higgs boson with TeV  
masses would be difficult at the presently operating LHC within MSSM, but would be possible within the NMSSM.

The single top-squark with lighter chargino production within natural SUSY has been studied 
in \cite{Hikasa:2015lma,Duan:2016vpp}. In \cite{Eckel:2014wta}, the authors have considered varieties 
of complex decay chains of lighter top-squark to find their signatures at the LHC.

{\it Generic top-squark experimental searches:} From the experimental side, top-squark SUSY searches have been performed lately by ATLAS and CMS experiments, studying mainly the decays into top quarks and neutralinos ($\tone\rightarrow  t\lsp$) and/or bottom quarks and charginos ($\tone\rightarrow b\chiapm$) \cite{cms:hadronic2017}, \cite{Asadi:2017qon}. Different analysis strategies are defined according to the number of leptons in the final state, to exploit the particular characteristics of the different signal topologies, aiming to high background discrimination. The dilepton channel  is the most relevant for our present analysis, the final topology studied for this channel corresponds to two opposite charged leptons (either electrons or muons), transverse energy imbalance (produced by the undetected neutralinos)  and jets from b-quarks. Transverse mass variables are very useful to reduce the dominant $t\bar t$ backgrounds \cite{MTvariables}. Masses of  $\tone$ $<$ 800 GeV (with $\mlsp <$ 360 GeV) have been excluded for the $\tone\rightarrow  t\lsp$ channel and  $m_{\tone}<$ 750 GeV (with $\mlsp<$ 320 GeV) for the $\tone\rightarrow b\chiapm$ channel, assuming 100$\%$ branching fractions in both cases \cite{cms:dilepton2017}. A search for pair production of top-squark with four or more jets and missing energy channel has been studied recently 
by ATLAS collaboration \cite{Aaboud:2017ayj}. No signal has been found, however, they put exclusion limits of the 
top-squark masses in the range between 450-950 GeV and 235-590 GeV. However, these exclusions depend on the mass 
differences with the lightest neutralino. The heavier top-squark has also been searched by ATLAS collaboration \cite{Aaboud:2017ejf}. 
Searches for  top-squark have also been performed by considering decays through flavor changing neutral current processes \cite{cms:hadronic2017} or considering resonant Higgs pair production \cite{Duan:2017zar}. The CMS collaborations has searched for 
lighter top-squark pair production in the all jets with missing energy channel in \cite{Sirunyan:2017wif}.

{\it  Associated and supersymmetric $ttH$ phenomenology:}

The standard model Higgs with top-quark associated production within the SM (taken into account CP-violation) has been 
studied in \cite{Ellis:2013yxa}, where the authors outline how to measure the Yukawa
couplings for the associated production of CP-even and CP-odd Higgses with top-quark pair and also the
the Higgs production with a single top-quark. They also point out how the spin-correlation between
the final top-quarks would explore more information about those Yukawa couplings. The non-standard lightest singlet
type Higgs together with pair of top quark has been studied recently in \cite{Badziak:2016tzl} without assuming
any correlation between the top-quark and gluon couplings to the Higgs boson. Taking into account the
perturbativity one can explore  higher values of the lighter top-squark with lighter singlet states.

The Supersymmetric version of $t \bar t H$ in a scenario with near mass degenerate lighter top-squark and
lightest higgsino has been studied  in \cite{Goncalves:2016tft}. In particular the authors have found the
optimal sensitivity for mass ranges with $\mtone$ - $\mlsp$ $\lsim$ $m_W$ and with $\mtone \lsim $380 GeV. Furthermore,
in the pure higgsino limit, where the production cross-section is expected to be largest from the couplings,
they outlined a strategy to extract the top--squark mixing angle from the angular distributions of the decay products.

To understand the Yukawa coupling of top-Higgs-top vertices, recently a jet-parton matching
(i.e., the detector objects matching with the underlying hard scattering) algorithm has been
developed using deep learning techniques \cite{Erdmann:2017hra}.

For the lighter top-squark as a NLSP (next-to-lightest superparticle) with very small mass difference to the   
lightest superparticle ($\mtone$- $\mlsp$) scenario, the discovery of the lighter top-squark is very difficult at the LHC.
However, adding an extra hard--jets \cite{Drees:2012dd} with the pair production, $\tone \tone j$ (similar Feynman 
diagrams of the $\tone \tone h_1$ that we considered in our analysis) -- termed as mono-jet searches would give
an extra handle to explore the enlarged $\tone$--$\lsp$ model spaces.\\

Very recently the observability of the light singlet-like NMSSM Higgs boson at
the LHC has been studied in \cite{Beskidt:2017dil}. This analysis is very close
to our focus, except for the lighter top-squark part. The authors use a novel
scanning technique to cover the whole NMSSM model parameter spaces, with
all the salient features the model posses and find several interesting benchmark
points to look for the non-standard Higgs boson signal at LHC. In our analysis,
we  cover this part including the lighter top-squark and moreover
for the most optimistic benchmark points, we are doing Signal and backgrounds analysis.

A comparative study for lighter top-squark phenomenology within MSSM and NMSSM has been 
performed in \cite{Cao:2012fz} from the SM-Higgs boson perspective. It is true that 
in MSSM, to be consistent with the lighter Higgs boson phenomenology, one needs 
some amounts fine-tunning in model spaces. However, in NMSSM because of having 
more parameters spaces, the fine-tunning could be evaded. The authors also briefly  
mention the lighter top-squark phenomenology that we focussed in our analysis.

{\it The associated and supersymmetric $ttH$ analysis at ATLAS:} The associated production of SM Higgs 
with top-quark pair followed by the decay of $h \to b \bar b$  has been reported in \cite{TicseTorres:2016kjh,Serkin:2017bqi}. 
From the experimental search the ATLAS collaboration \cite{TheATLAScollaboration:2016tsz,ATLAS:2017tmd} has been 
searching for the direct pair production of lighter top-squark pairs in the di-lepton channel, considering 
all possible decay modes of the top-squarks, namely the two body, three-body, four-body. There is no excess 
of signal of top-squark over background, however the exclusion limits became stronger. 

{\it The associated and supersymmetric $ttH$ analysis at CMS:} 

They have also been looking for the SM $t \bar t h$ 
production followed by the decays of $h \to b \bar b$ \cite{CMS:2016qwm}.  The top-squark pair production 
in the opposite sign dilepton channel has been studied recently by CMS collaboration\cite{CMS:2016kcq}.
The lighter top-squark (and also lighter bottom squark) has been searched in the two body decay channel and
has been reported recently in \cite{Sirunyan:2017kiw}. The CMS collaboration has also been looking the associated production of lighter (heavier) top-squark pair together with a 
SM Higgs boson (and Z-boson) \cite{CMS:2014bla}. 

It is clear from all the existing analysis to date that the top-squark with low masses is still one of the candidates to discover supersymmetry. Lighter top-squark together with the non-standard  
Higgs is an extra handle to establish the beyond SM phenomenon more profoundly.  
We are considering this possibility in our analysis here.

The plan of this paper is as follows. In the next section we  briefly describe the NMSSM model.
In Sec.2, we randomly vary the NMSSM model parameters and identify the allowed parameter space
consistent with up-to-date theoretical, phenomenological and experimental constraints.
For the allowed model spaces, we then estimate the number of SUSY associated non-standard Higgs production with lighter top-squark events, $pp \to \tone \hone \tone$ with the decay 
channel of $\tone \to b \chiapm$, $\chiapm \to W \chia$ and $h_1 \to b \bar b$ and identify 
few high event-rated benchmark points for LHC energy with 13 TeV and a future 33 TeV proton collider, to carry out 
the phenomenological analysis in Sec.3. In doing so, we estimate all the reducible and irreducible 
SM backgrounds for the signal channel under consideration. In Sec.4, we carry out detector 
level analysis to isolate the non-standard Higgs boson and top-squark signals. We summarize and conclude our findings in Sec.5.

\section{The NMSSM models}
\label{sec:nmssm}

The NMSSM model has been described in many reviews \cite{nmssm,Drees:1988fc,Franke:1995tc,Maniatis:2009re,review}. 
Here, however let us mention the relevant part following \cite{review} for our analysis. 
The NMSSM contains one additional gauge singlet chiral superfield, $\widehat{S}$ compared to MSSM superfields.  
The Higgs superpotential $W_\mathrm{Higgs}$ is written as
\beq\label{higgssuper}
W_\mathrm{Higgs} = (\mu + \lambda \widehat{S})\,\widehat{H}_u \cdot
\widehat{H}_d + \xi_F \widehat{S} + \frac{1}{2} \mu' \widehat{S}^2 +
\frac{\kappa}{3} \widehat{S}^3,
\eeq
where $\lambda$, $\kappa$ are dimensionless Yukawa couplings, $\mu$, $\mu'$ are the 
supersymmteric mass terms with mass 
dimension one, and $\xi_F$ is the supersymmetric tadpole term, with mass-dimension two. 
Assuming R-parity and CP-conservation (scenarios violating this discrete symmetries  have been  
studied in \cite{JeanLouis:2009du} and \cite{Goodsell:2016udb}) the 
soft--supersymmetry breaking terms, ${\cal L}_\mathrm{soft}$ are the following:

\bea\label{2.5e}
-{\cal L}_\mathrm{soft} &=&
m_{H_u}^2 | H_u |^2 + m_{H_d}^2 | H_d |^2
+ m_{S}^2 | S |^2+m_Q^2|Q^2| + m_U^2|U_R^2| \nn \\
&&+m_D^2|D_R^2| +m_L^2|L^2| +m_E^2|E_R^2|
\nn \\
&&+ (h_u A_u\; Q \cdot H_u\; U_R^c - h_d A_d\; Q \cdot H_d\; D_R^c
- h_{e} A_{e}\; L \cdot H_d\; E_R^c\nn \\ &&
+\lambda A_\lambda\, H_u \cdot H_d\; S + \frac{1}{3} \kappa A_\kappa\,
S^3 + m_3^2\, H_u \cdot H_d + \frac{1}{2}m_{S}'^2\, S^2 + \xi_S\, S
+ \mathrm{h.c.}) \; ,
\eea

where $m_{H_u}, m_{H_d}$ are the Higgs up and down-type soft mass terms, respectively. The singlet soft-mass  
parameter is $m_{S}$. The left-handed quark (lepton)  doublet mass is $m_Q$ ($m_L$), while $m_U$ and $m_D$ ($m_E$) are the 
right-handed singlet mass term for up-type quark and down-type quark (lepton) superfields. The 
$h_{u,d,e}$ ($A_{u,d,e}$) is the Yukawa couplings (tri-linear soft mass parameters) for up, down type quarks and down type lepton, respectively. 
The associated soft supersymmteric 
mass terms with mass dimension one $m_3$, $m_{S}'$ and with mass dimension three $\xi_S$ have 
to be of the order of the weak or SUSY breaking scale.

All these above terms are generically non-vanishing, however the 
scale invariance leads to a simplified version with $\mu=\mu'=\xi_F = 0$, together with the parameters $m_3^2$, $m_{S}'^2$ and $\xi_S$ in 
(\ref{2.5e}) also set to zero.
Thus, the superpotential takes the following form 

\beq\label{scaleinv}
W_\mathrm{scale-invariant} = \lambda \widehat{S}\,\widehat{H}_u \cdot
\widehat{H}_d + \frac{\kappa}{3} \widehat{S}^3.
\eeq

While the singlet superfield  
$\widehat{S}$ gets a VEV at the SUSY breaking scales, an 
effective $\mu$-term of the order of weak scale is the following the simple form 
\beq\label{2.7e}
\mu_\mathrm{eff} = \lambda s\; . 
\eeq

The Higgs potential is obtained from the supersymmetric gauge interactions, 
the $F$-term and the soft supersymmetry breaking terms:  
\bea
V_\mathrm{Higgs} & = & \left|\lambda \left(H_u^+ H_d^- - H_u^0
H_d^0\right) + \kappa S^2 + \mu' S +\xi_F\right|^2 \nn \\
&&+\left(m_{H_u}^2 + \left|\mu + \lambda S\right|^2\right)
\left(\left|H_u^0\right|^2 + \left|H_u^+\right|^2\right)
+\left(m_{H_d}^2 + \left|\mu + \lambda S\right|^2\right)
\left(\left|H_d^0\right|^2 + \left|H_d^-\right|^2\right) \nn \\
&&+\frac{g_1^2+g_2^2}{8}\left(\left|H_u^0\right|^2 +
\left|H_u^+\right|^2 - \left|H_d^0\right|^2 -
\left|H_d^-\right|^2\right)^2
+\frac{g_2^2}{2}\left|H_u^+ H_d^{0*} + H_u^0 H_d^{-*}\right|^2\nn \\
&&+m_{S}^2 |S|^2
+\big( \lambda A_\lambda \left(H_u^+ H_d^- - H_u^0 H_d^0\right) S +
\frac{1}{3} \kappa A_\kappa\, S^3 + m_3^2 \left(H_u^+ H_d^- - H_u^0
H_d^0\right) \nn \\
&& +\frac{1}{2} m_{S}'^2\, S^2 + \xi_S\, S + \mathrm{h.c.}\big)
\label{2.9e}
\eea

where $g_1$ and $g_2$ are $U(1)_Y$ and $SU(2)$ gauge couplings, respectively.

The full scalar potential (\ref{2.9e}) has been expanded around the real neutral VEVs $v_u$, $v_d$ and $s$ (up, down and singlet VEV respectively) and the physical neutral Higgs fields  are the following forms:
\beq\label{2.10e}
H_u^0 = v_u + \frac{H_{uR} + iH_{uI}}{\sqrt{2}} , \quad
H_d^0 = v_d + \frac{H_{dR} + iH_{dI}}{\sqrt{2}} , \quad
S = s + \frac{S_R + iS_I}{\sqrt{2}}\; ;
\eeq
where the CP-even part is labeled with index $R$, while index $I$ is used for the CP-odd states. The VEVs have to be obtained from the minima of
\bea
V_\mathrm{Higgs} & = & \left(-\lambda v_u v_d + \kappa s^2 + \mu' s
+\xi_F\right)^2 +\frac{g_1^2+g_2^2}{8}\left(v_u^2 - v_d^2\right)^2
\nn \\
&&+\left(m_{H_u}^2 + \left(\mu + \lambda s\right)^2\right) v_u^2
+\left(m_{H_d}^2 + \left(\mu + \lambda s\right)^2\right) v_d^2
\nn\\
&&+m_{S}^2\, s^2 -2 \lambda A_\lambda\, v_u v_d s +  \frac{2}{3}
\kappa A_\kappa\, s^3 - 2m_3^2\, v_u v_d + m_{S}'^2\, s^2 +
2\xi_S\, s \;,
\label{2.11e}
\eea

The minimization of (\ref{2.11e}) with respect the three VEVs and the 
proper radiative electroweak symmetry breaking ( for generating the correct
$Z$-boson mass) leads to the following input parameters:  

\beq\label{2.17e}
\lambda,\ \kappa,\ A_\lambda,\ A_\kappa,\ \tan\beta,\
\mu_\mathrm{eff},
\eeq
to which one has to add the (in the convention $\mu = 0$) five parameters of the NMSSM 
\beq\label{2.18e}
m_3^2,\ \mu',\ m_{S}'^2,\ \xi_F\ \mathrm{and}\ \xi_S\; .
\eeq

The full scalar potential (\ref{2.9e}) has to be expanded around the real neutral VEVs $v_u$, $v_d$ and $s$,  
as in (\ref{2.10e}), to get the tree-level Higgs mass matrices. The matrix elements of the $3 \times 3$ CP-even  
mass matrix ${\cal M}_S^2$ are conveniently written in the basis $(H_{dR}, H_{uR}, S_R)$ after the elimination 
of $m_{H_d}^2$, $m_{H_u}^2$ and $m_{S}^2$. 
Likewise, the $3 \times 3$ CP-odd mass matrix ${\cal M'}_P^2$ are written in the basis of the imaginary part of the down, up and singlet $(H_{dI}, H_{uI}, S_I)$ fields. After dropping the Goldstone mode,  
the remaining $2 \times 2$ CP-odd mass matrices has the following input parameters: the doublet ($M_A$) 
and singlet component ($M_P$) mass parameters together with the $\mu_{eff}$. 
In our analysis we consider a general phenomenological NMSSM, which is different from the $Z_3$ invariant NMSSM. 
However, imposing the $m_3^2 = m_{S}'^2 = \xi_S = \mu = \mu' =\xi_F = 0$ constraints in the general 
phenomenological NMSSM, one can recover the $Z_3$ invariant NMSSM. 

The mass matrix of the top-squark in the left-right interactions basis $(\tilde{t}_L,\tilde{t}_R)$ is given by \cite{Baglio:2015noa}: 
\beq
{\cal M}^2_{\tilde{t}} = \left( \begin{array}{cc} m_{LL}^2 & m_{LR}^2
    \\ m_{RL}^2 & m_{RR}^2 \end{array} \right) \;,
\eeq
with
\beq
m_{LL}^2 = m_{\tilde{Q}}^2 + m_t^2 + M_Z^2 \cos 2\beta \left( \frac{1}{2} -
\frac{2}{3} \sin^2 \theta_W \right)
\eeq

\beq
m_{RR}^2 = m_{\tilde{t}_R}^2 + m_t^2 + \frac{2}{3 }M_Z^2 \cos 2\beta \sin^2
\theta_W 
\eeq

\beq
m_{LR}^2 = m_{RL}^2 = m_t (A_t - \mu \cot\beta) \;,
\eeq

where $m_{\tilde{Q}}$ ( $m_{\tilde{t}_R}$) is the common left (right) -handed soft SUSY breaking mass,  
$A_t$ is the tri-linear soft SUSY breaking mass parameter, $\mu$ is the higgsino mass parameter, 
$m_t$ and $M_Z$ are the top and Z-boson mass respectively and $\theta_W$ is the Weinberg angle.
The top-squark mass matrix is diagonalized by

\beq
{\cal R}^{\tilde{t}} = \left( \begin{array}{cc} \cos
    \theta_{\tilde{t}} & \sin \theta_{\tilde{t}} \\ -\sin
    \theta_{\tilde{t}} & \cos \theta_{\tilde{t}}  \end{array} \right)
\eeq

leading to two top-squark mass eigenstates $\tilde{t}_i$ ($i=1,2$) as

\beq
\tilde{t}_i =  {\cal R}^{\tilde{t}}_{is} \tilde{t}_s \;,
\eeq

where $s=L,R$ and by convention $m_{\tilde{t}_1} <
m_{\tilde{t}_2}$. The top-squark mixing angle $\theta_{\tilde{t}}$ and the 
mass-eigenstates are the following form:

\beq
\tan \theta_{\tilde{t}} = \frac{2
  m_{LR}^2}{m_{LL}^2-m_{RR}^2-\sqrt{(m_{LL}^2-m_{RR}^2)^2 + 4
    m_{LR}^4}},
\eeq

and

\beq
m_{\tilde{t}_{1,2}}^2 = \frac{1}{2} \left[ m_{LL}^2 + m_{RR}^2
\mp \sqrt{(m_{LL}^2-m_{RR}^2)^2+4 m_{LR}^4} \right] \;.
\eeq

The mixing and hence, the diagonalization, leads to the the lighter top-squark to be very light around the 
mass of the SM top-quark. This particular mass ranges (with lightest neutralino less than the weak-gauge bosons called the compressed scenario in the literature) are very challenging from the experimental perspective as 
several decay modes are competing and yet to be excluded with high enough confidence level. In the next sections, 
we perform the NMSSM model parameter scanning and delimit the low mass top-squark satisfying all 
the available theoretical, phenomenological and physical constraints.

\section{The NMSSM parameter spaces}
\label{sec:NMSSMparam}

In our model space scanning we used the package \texttt{NMSSMTools~5.0.1}~\cite{Ellwanger:2004xm} to obtain the superparticle masses, decay branching ratios and various low energy observables. We randomly scanned approximately 
$10^7$ points. The varied parameters and their ranges are tabulated in Table \ref{tab:param}.

\begin{table}[h]
\caption{The minimum and maximum values of varied NMSSM parameters. 
The following parameters remain fixed: $M_3$ = 1900.0 GeV (this allows the gluino mass $\mgl$ to be above the mass limits from recent LHC-run2); $m_{\tilde \ell}$ = 300.0 GeV
(for all three generation as well as left and right state) and $A_{\tau}$=$A_e$=$A_{\mu}$ = 1500.0 GeV. 
These particular choices have been chosen such that the slepton masses naturally satisfy their experimental  
bounds. Here $M_A$ ($M_P$) is the Doublet(Singlet) component of the CP-odd Higgs mass matrices. All masses and mass parameters in our 
analysis are in GeV. }
\label{tab:param}
\centering
\begin{tabular}{c|cc}
\hline\hline
Parameters&Min&Max\\
\hline
$\lambda$&0.001& 0.7\\
$\kappa$& 0.001& 0.7\\
$A_{\lambda}$&100.0& 2500.0\\
$A_{\kappa}$& -2500.0& 100.0\\
$\tanb$&1.5 & 60.0 \\
$\mu_{eff}$& 100.0& 500.0\\
$M_1$& 50.0& 400.0\\
$M_2$& 50.0& 500.0\\
$\msql$& 300.0& 1500.0\\
$A_t$=$A_b$& -4000.0& 1000.0\\
$M_A$& 100.0& 500.0\\
$M_P$& 100.0& 3000.0\\
\hline
\hline
\end{tabular}
\end{table}

\begin{figure}[ht!]
\begin{center}
\includegraphics[scale=0.145]{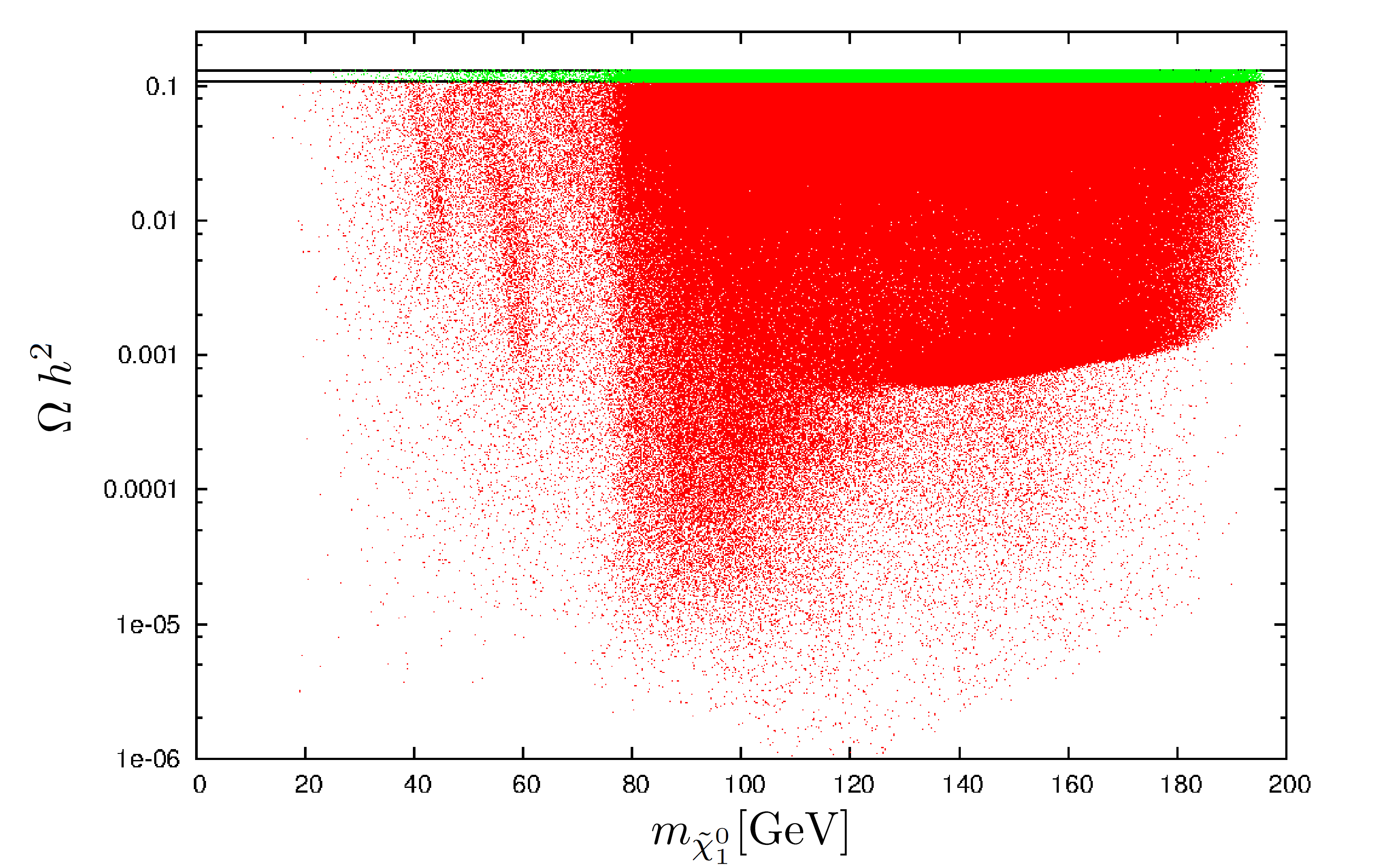}

\includegraphics[scale=0.14]{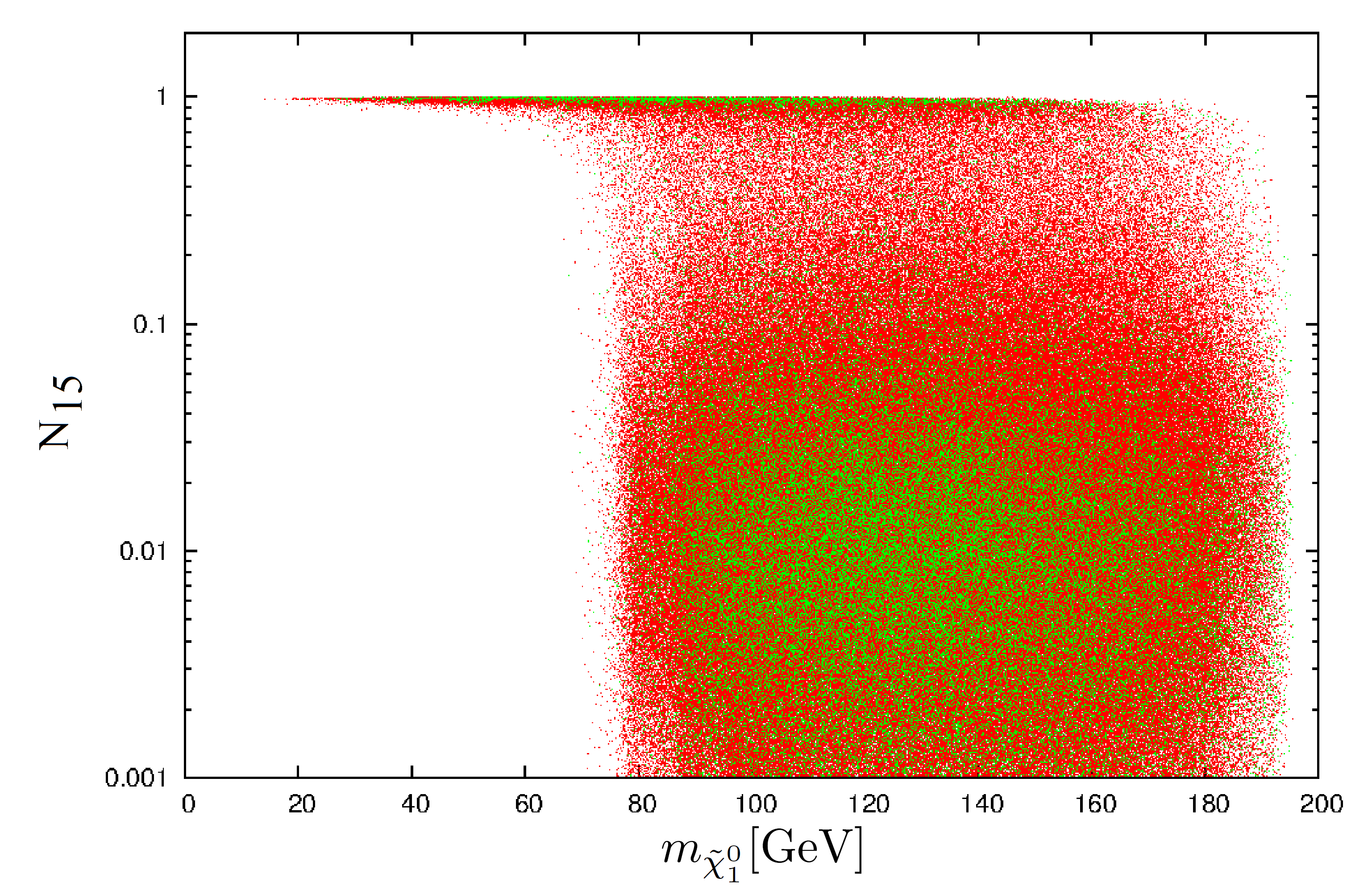}
\caption{Upper panel: The Dark matter relic density $\Omega_{\lsp} h^2$ as a function
of lightest neutralino masses $\mlsp$ within the standard cosmological model. The two lines
represent the upper ($\Omega_{\lsp} h^2$=0.131) and lower ($\Omega_{\lsp} h^2$=0.107)
bounds from the Planck measurements \cite{Planck:2015xua}. The first deep around 45 GeV
is due to the Z-boson exchange (annihilation diagram) and the second around 63 GeV,
from the Higgs-boson ($h_2$-SM) exchange annihilation within the NMSSM parameter spaces.
The green points within the strips satisfy the direct and indirect dark matter searches results. 
Lower panel: The singlet component $N_{15}$ of the 
lightest neutralino $\lsp$ consistent with the dark matter relic density $\Omega_{\lsp} h^2$ and various other phenomenological constraints.
}
\label{dm}
\end{center}
\end{figure}

For each randomly generated parameter spaces, we invoke the following constraints:

\begin{description}
\item[Perturbative bounds:] All points must satisfy $\lambda^2 + \kappa^2 \lsim (0.7)^2$ \cite{Zheng:2014loa}.  
This values ensures that the NMSSM or the minimal $\lambda$-SUSY model spaces remain perturbative up to  
the GUT-scale (without invoking any new fields in between the weak-scale and the GUT-scale).   
\item[Dark Matter relic density:] We considered the standard cosmological scenario with the lightest neutralino  
as the WIMP (weakly interacting massive particle) dark matter candidate within the standard cosmological model scenario. 
We demanded the relic density should be within the range 0.107 $<$ $\Omega_{\lsp} h^2$ $<$ 0.131, consistent with 
the Planck measurement \cite{Planck:2015xua}. The estimated relic density $\Omega_{\lsp} h^2$ as a function of the $m_{\lsp}$ has been shown in the upper panel of Fig.\ref{dm} with $\Omega_{\lsp} h^2$ $<$ 0.131
(by red-points). The green-marked points within the upper and lower strips are consistent with the direct and indirect detection bounds.
To estimate this $\Omega_{\lsp} h^2$ limits, the \texttt{NMSSMTools 5.0.1} is interfaced  
with \texttt{micrOMEGAs v4.3} \cite{micr43,Belanger:2013oya}. 

\item[Higgs bounds:] We consider the intermediate Higgs boson ($h_2$) is the SM-like Higgs boson with masses in the ranges $ 125.09\ \mathrm{GeV} < m_{h_2} < 128.09\ \mathrm{GeV}$, taking into consideration the 3 GeV error in the theoretical estimates, and from the coupling ratios 
and signal strength measurements from LHC-run1 ATLAS and CMS combined studies \cite{Khachatryan:2016vau}.  
The allowed coupling ratios and the measured signal strengths considered in our analysis has been tabulated in Table \ref{tab:coupsigma}. The invisible branching ratio on the SM-like Higgs 
boson has also been invoked: $BR(h_{SM} \rightarrow invisible)$ $\lsim$ 0.25 \cite{Khachatryan:2016whc,Aad:2015pla}  
{\footnote{The invisible decay of SM-like Higgs boson within the NMSSM has been studied  
in \cite{Butter:2015fqa}.}}. Furthermore, we required the masses of the charged Higgs boson to be $m_{\hpm} > 80.0\ \mathrm{GeV}$. 

\item[LEP bounds:] Direct SUSY searches of the LEP experiments had set bounds on the lighter chargino with $m_{\chiapm} >$ 103.5 GeV. On the other hand, the ALEPH experiment puts a strong constraint on the $Z$ invisible width, to satisfy $\Gamma_Z^\mathrm{inv}$ $< 2~$ MeV at 95\% C.L. ~\cite{ALEPH:2005ab}.   
When the decay channel $Z \to \lsp\lsp$ opens, this width may exceed the experimental value   
{\footnote{A light Higgs boson would significantly affect the anomalous magnetic moment of muon: $a_\mu=(g_\mu-2)/2$,    
whose most accurate measurement comes from the E821 experiment \cite{Bennett:2004pv}.     
However, since the measurements has large theoretical uncertainties, we have not consider this constraints in our model space scanning.}}. 

\item[B physics bounds:] We consider the flavor constraints coming from the rare decays of $B$-meson, such 
as $B_s\to \mu^+\mu^-$, $B^+ \to \tau^{+} \nu$, and $B_s\to X_s\gamma$. In our numerical scan, we set the recent  
experimental results at $95\%$ C.L.: $1.7 \times 10^{-9} < \br(B_s \to \mu^{+} \mu^{-}) < 4.5 \times 10^{-9}$~\cite{Amhis:2014hma},
$0.85 \times 10^{-4} < \br(B^+ \to \tau^{+} \nu) < 2.89 \times 10^{-4}$~\cite{Lees:2012ju}, and
$2.99 \times 10^{-4} < \br(B_s\to X_s\gamma) < 3.87 \times 10^{-4}$~\cite{Amhis:2014hma}.

\item[Sparticle masses:] The superparticle masses should satisfy following \cite{Aaboud:2016tnv}: 

$\mgl$ $\gsim$ 1700.0 GeV,
$\mtone$ $\gsim$ 95.0 GeV, $\mbone$ $\gsim$ 325.0 GeV, 
$\msql$ $\gsim$ 600.0 GeV, $\mlepl$ $\gsim$ 100.0 GeV, 
$\mnulepl$ $\gsim$ 90.0 GeV and $\mtauone$ $\gsim$ 87.0 GeV.

\end{description}

\begin{table}[t]
\caption{The couplings ($\kappa$) and signal strength ($\mu$) have been allowed within 2$\sigma$ ranges (except for $\kappa_W$) from the combined ATLAS and CMS measurements 
\cite{Khachatryan:2016vau}.
 }
\label{tab:coupsigma}
\centering
\begin{tabular}{c|cc|c|cc}
\hline\hline
Parameters&Min&Max & Parameters&Min&Max \\
\hline
$\kappa_W$&0.81&0.99 & $\mu_{VBF}^{\tau \tau}$&0.50&2.10 \\
$\kappa_{t}$&0.99&1.89 & $\mu_{ggF}^{\tau \tau}$&-0.20&2.20\\
$|\kappa_{\gamma}|$&0.72&1.10 & $\mu_{VH}^{bb}$&0.00&2.00\\
$|\kappa_{g}|$&0.61&1.07 & $\mu_{ttH}^{bb}$&-0.90&3.10\\
$|\kappa_{\tau}|$&0.65&1.11 & $\mu_{VBF}^{WW}$&0.40&2.00\\
$|\kappa_{b}|$&0.25&0.89 &$\mu_{ggF}^{ZZ}$&0.51&1.81\\
$Br(h_{SM} \to inv.)$& & 0.25 &$\mu_{VBF}^{\gamma\gamma}$&0.30&2.30\\
&& &$\mu_{ggF}^{\gamma\gamma}$&0.66&1.56\\
\hline
\hline
\end{tabular}
\end{table}

If the randomly generated NMSSM model space satisfy all the above physics constraints, we consider them for further
phenomenological studies. The relic density $\Omega_{\lsp} h^2$ as a function of the lightest neutralino masses 
is shown in the upper panel of Fig.\ref{dm}. All the points satisfy the lower and upper bounds of 0.107 and 0.131, respectively, coming from the recent Planck measurements \cite{Planck:2015xua}. Within this strip, the green points satisfy the direct and indirect Dark Matter searches.
The lightest neutralino co-annihilation would occur via the $Z$-boson (SM-Higgs boson, $h_2$) exchange diagram, which shows a dip around $M_Z/2$ ($m_{h_2}/2$), i.e, 45 (63) GeV.  

In the lower panel of Fig.\ref{dm}, we estimate the singlet composition of the lightest neutralino $N_{15}$ as function of the lightest neutralino masses. Constraints coming from the sparticles masses, B-physics and other phenomenological limits are satisfied,
together with the relic density 
constraints. 
After passing this criterion we invoked the constraints from  
the SM-Higgs boson, namely the intermediate Higgs boson $h_2$, 
satisfying observed masses, couplings modifiers and signal strength following Table.\ref{tab:coupsigma} called the $h2$-SM scenario.

\begin{figure}[ht!]
\label{fig:massbr}
\begin{center}
\includegraphics[scale=0.5]{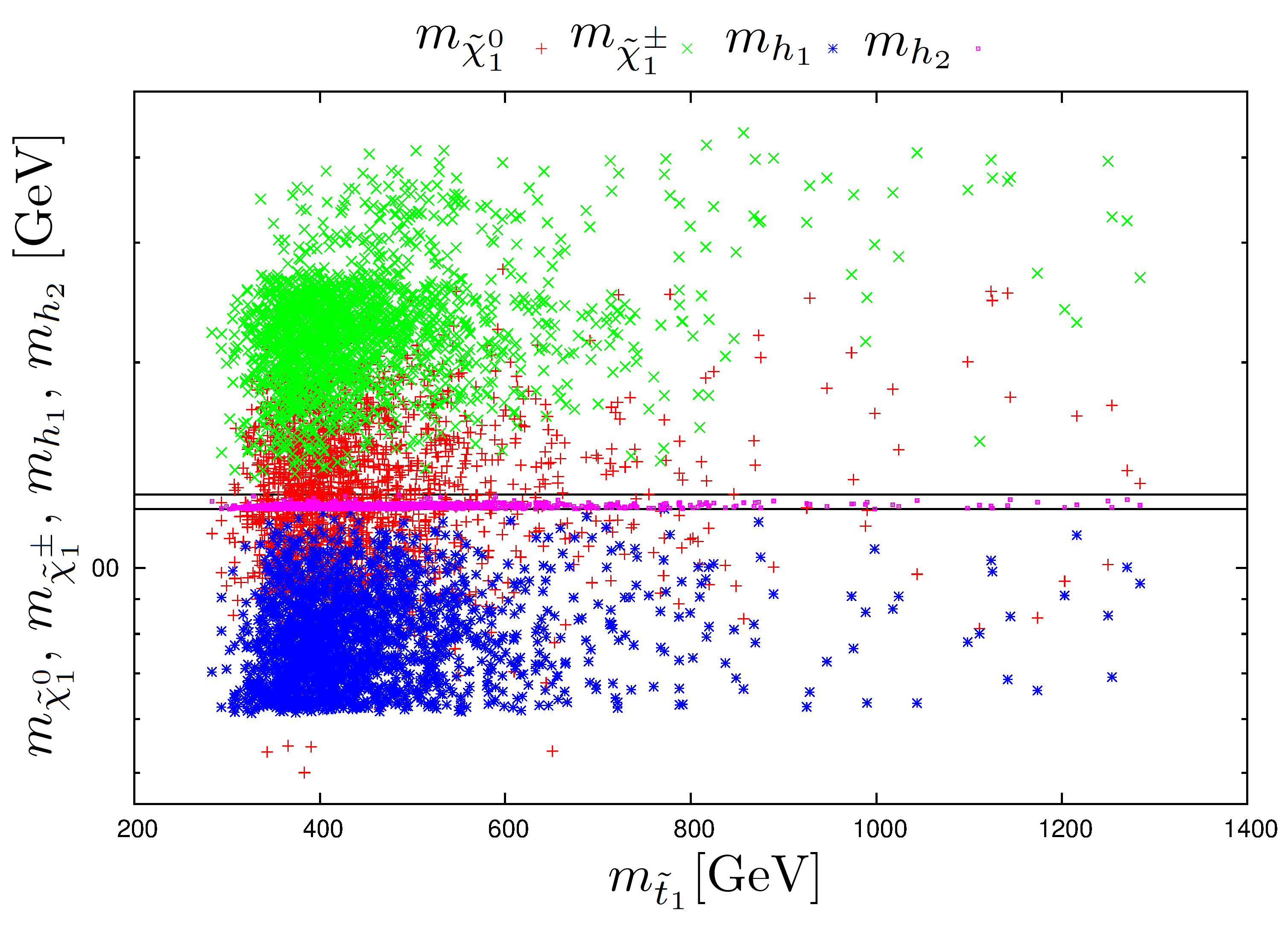}

\includegraphics[scale=0.5]{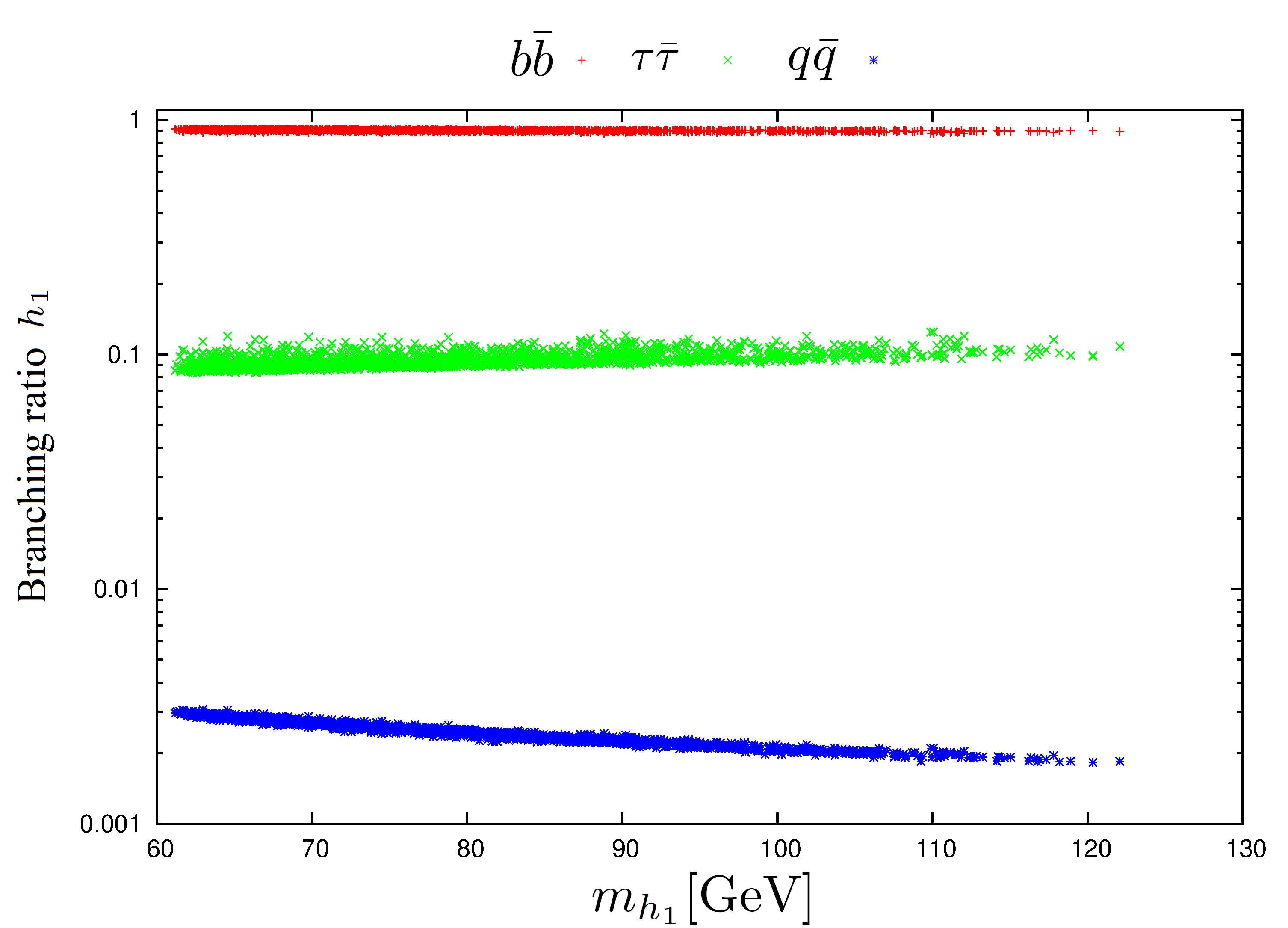}
\caption{Upper panel: The masses of $h_1$, $h_2$ $\chia$ and $\chiapm$ as a function of 
lighter top-squark mass $\tone$ for the $h_2$-SM scenario and consistent with all other constraints (see details in the text).
The masses of the $h_2$-SM, consistent with coupling ratios and signal strength following Table \ref{tab:coupsigma}, is shown 
within the horizontal lines for allowed ranges: [122.1-128.1] GeV. Lower panel: Branching ratios of $h_1$ into 
the $b \bar b$, $\tau \bar \tau$, and $q \bar q$, where $q$=$guds$.  It is clear that in a large
region of allowed parameter space, the branching ratio of $h_1 \to b \bar b$ is above 90\%.
}
\end{center}
\end{figure}

In our analysis, it is interesting to look for the lighter top-squark and non-standard Higgs boson in this $h2$-SM model space.   
In the upper panel of Fig.\ref{fig:massbr}, we show the masses of lighter non-standard Higgs boson $m_{h_1}$, $m_{h_2}$, lightest neutralino, lightest chargino 
as a function of the lighter top-squark masses in the $h2$-SM scenario. 
In the lower panel of Fig.\ref{fig:massbr}, we show the branching ratio of the $h2$-SM in the three most dominating channels.
For all the points in Fig.\ref{fig:massbr}, we estimated the event rates for the 
associated production of $\tone \tone h_1$ for the LHC energy under consideration. 
We  describe these details in the following section.

 For our allowed parameter spaces with $h_2$ as SM Higgs type, the masses of the lighter top-squark extend from 300 to 1300 GeV. The higher values of the lighter top-squark 
are always permissible with the large input of the common top-squark masses. The low-mass spectrum is mostly controlled from the mixing matrices (see sec.\ref{sec:nmssm}).  

\begin{figure}[ht!]
\label{topcharbr}
\begin{center}
\includegraphics[scale=0.5]{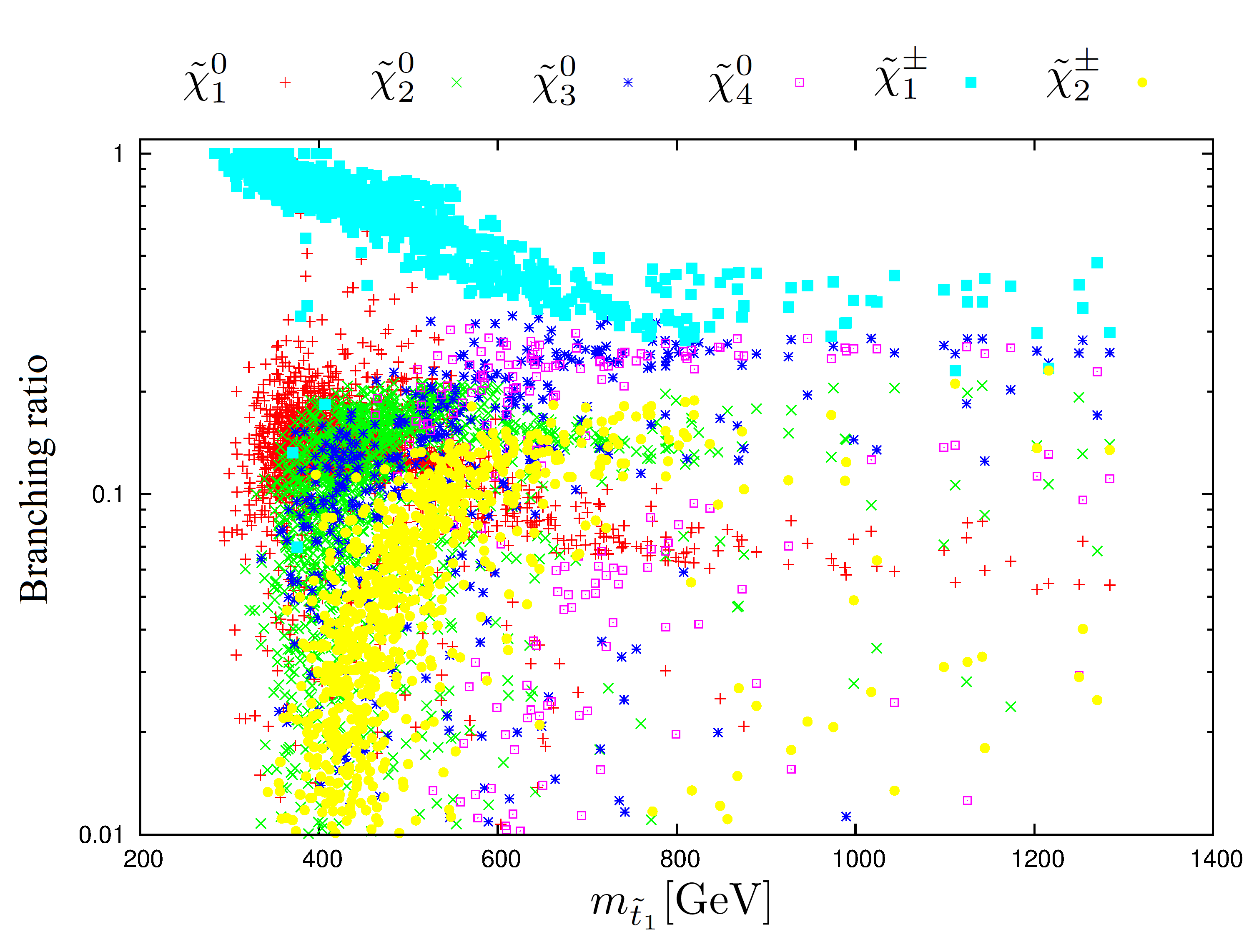}

\includegraphics[scale=0.5]{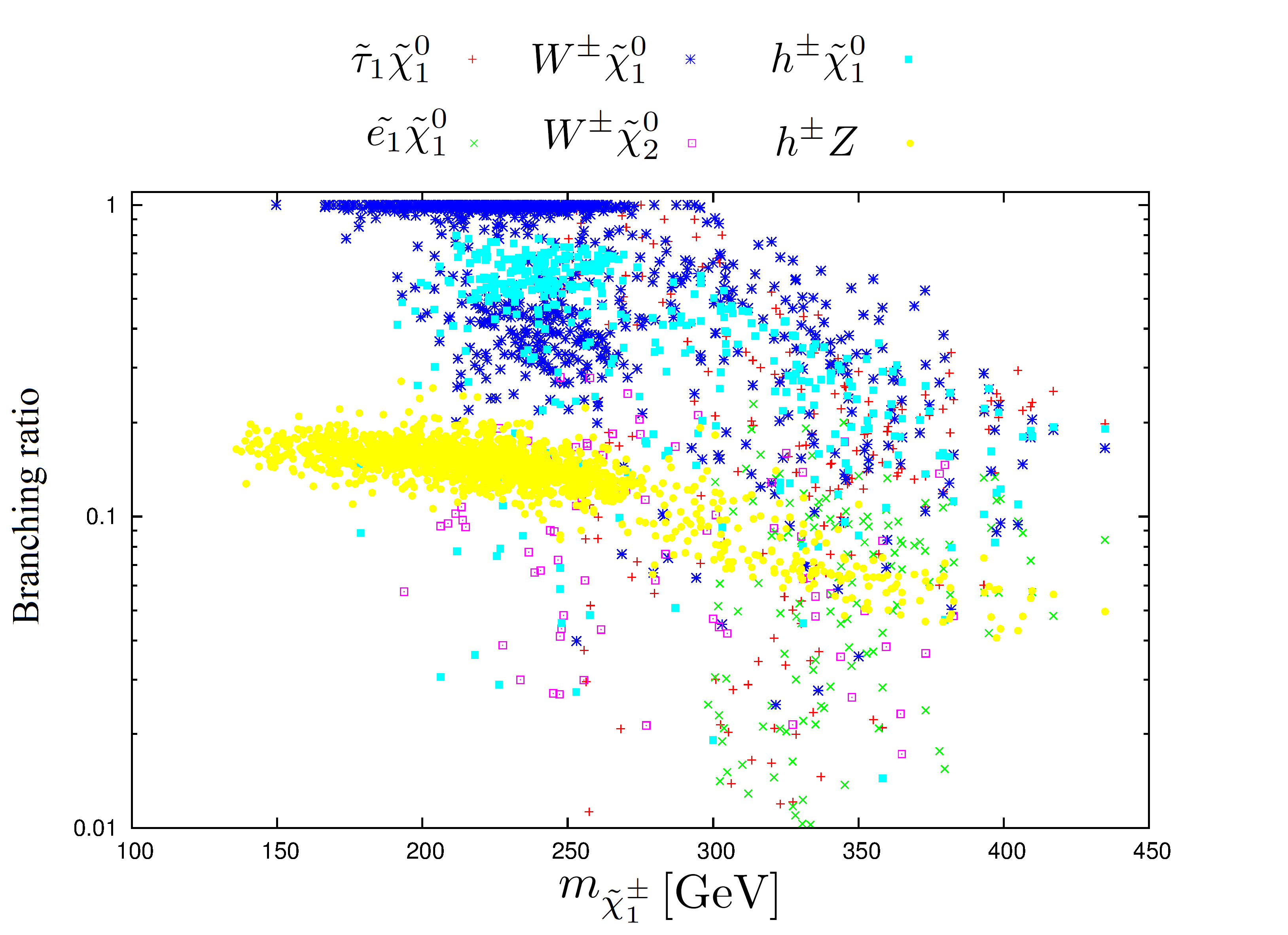}
\caption{Upper panel: Branching ratio of $\tone$ for the $h_2$-SM scenario and consistent with all other constraints. 
It is clear that the $b \chiapm$ is  dominating in the low mass regions. Lower panel: Branching ratio of $\chiapm$ 
and it seems it  mostly decays into $W{^\pm} \chia$. Because of the leptonic universality,  BR($\tilde{e_1}\tilde{\chi}_1^0$) = BR($\tilde{\mu_1}\tilde{\chi}_1^0$), so we show only the $\tilde{e_1}$ channel.
}
\end{center}
\end{figure}

\begin{figure}[ht!]
\label{fig:sigevent}
\begin{center}
\includegraphics[scale=0.16]{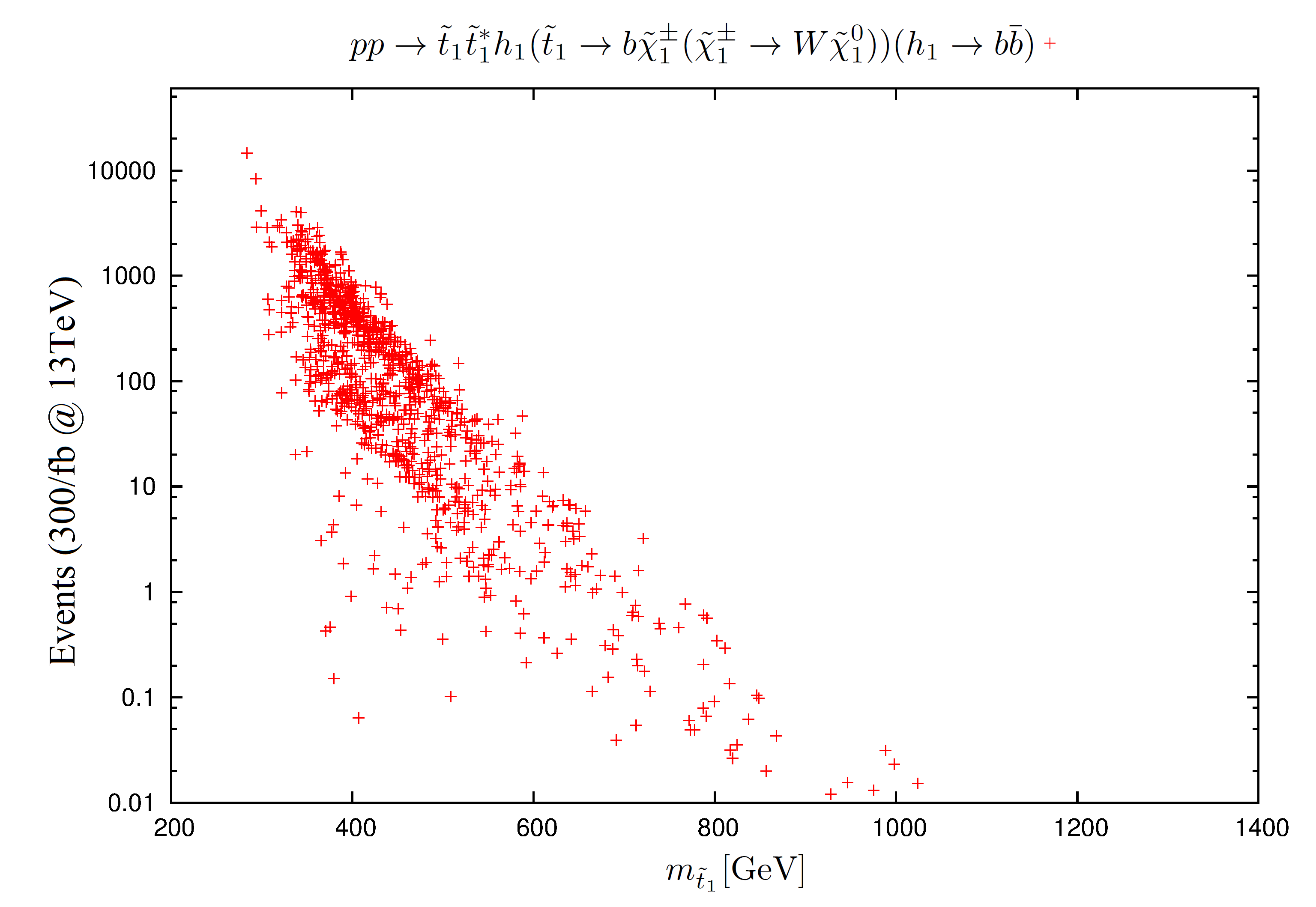}

\includegraphics[scale=0.5]{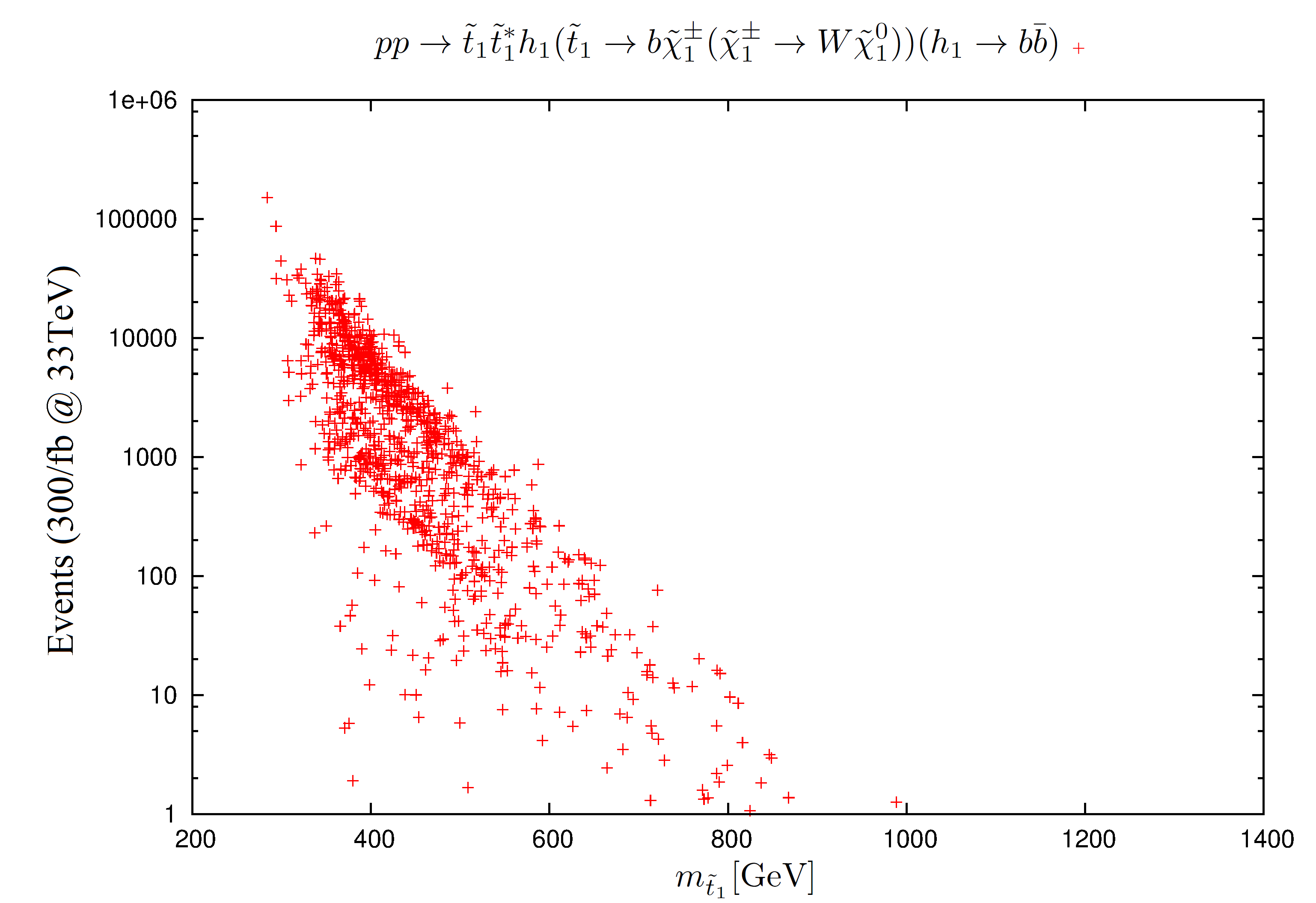}
\caption{Number of Signal events with an integrated luminosity of 300 fb$^{-1}$  at  
a center of mass energy of $\sqrt{s}=$13 (33 ) TeV in the upper (lower) panel.
 We consider both the 
lighter top-squarks are decaying as $\tone \rightarrow b \chiapm \rightarrow b W \chia$. 
The branching ratios of the $W$ bosons decaying into lepton have not been considered here, 
however the corresponding factor is multiplied in the event flow analysis. 
} 
\end{center}
\end{figure}

Depending upon the masses of top-squark and other sparticles spectra, the decay patterns are modified. This will guide us to choose the most optimistic decay cascade to look for the non-standard Higgs boson  
together with the lighter top-squark. We show the branching ratio of $\tone$ in the  
upper panel of Fig. \ref{topcharbr}. As it is clear from the figure, for the low mass of the top-squark, the dominating decay mode is $\tone \ra b \chiapm$ and could be more than 95\%. In few of the allowed  parameters spaces, if the lightest neutralino is higgsino type and if it is kinematically allowed then  
$\tone \ra t \chia$ could also be as large as 90\%. For larger masses, various other decay modes open up and compete with each other. As we are interested in exploring the low mass top-squark scenario, thus,  
we consider the $\tone \ra b \chiapm$ channel. In the next level of cascade, the $\chiapm$ undergoes decay, and we show the corresponding decays in the lower panel of Fig. \ref{topcharbr}. It seems for the low masses of chargino, the dominant decay is $\chiapm \ra W^{\pm} \chia$. In some parameter spaces, for the  
low mass charged Higgses, the $\hpm Z$ ($\hpm \chia$) channel would open up with somewhat appreciable branching ratios. We are not exploring the charged Higgs sector in this analysis, thus we are considering the 
$\chiapm \ra W^{\pm} \chia$ decay mode. We consider both the top-squarks decay into identical decay cascades. At the end, the two $W^{\pm}$ bosons are allowed to decay in the leptonic ($\ell$ = $e$ and $\mu$)  modes.  
The number of signal events with the decay cascades is shown in the upper(lower) panel of Fig. \ref{fig:sigevent} for proton collisions at 13(33) TeV. We have selected the best three large signal events benchmark points to do  
our numerical simulation in the following section.

\clearpage

\section{Numerical Analysis}
\label{sec:numerical}

\begin{figure}[ht!]
\centering
\includegraphics[scale=0.51]{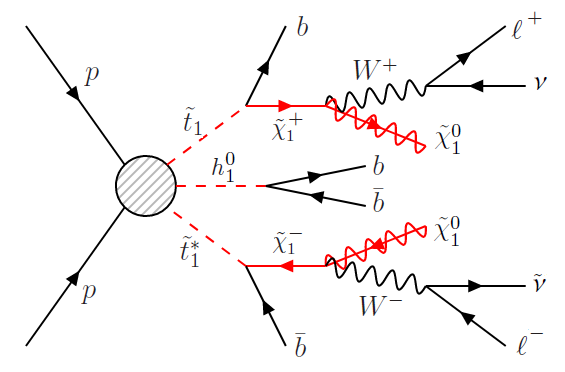}
\caption{Effective Feynman diagram illustrating the $pp\rightarrow \tone^* \tone \hone$ production with the final topology that we study.}
\label{fig:feynman}
\end{figure}

In the allowed NMSSM model parameter space, we find that the lighter top-squark mainly decays  
into $\tone \rightarrow b \chiapm$ (Fig.\ref{topcharbr} upper panel), and subsequently, the lighter chargino mainly decays into $\chiapm  \rightarrow W \chia$ (Fig. 2 lower panel). 
On the other hand, the non-Standard Higgs boson mainly decays into a $b \bar b$  pair (Fig.\ref{fig:massbr} lower panel). We focus on the topology where the two $W$-bosons decay into a pair of leptons (electrons or muons). We end up with  the final state $b\bar{b}b\bar{b}\ell\ell\nu\nu\tilde{\chi}_1^0\tilde{\chi}_1^0$ (or $4 b-\text{jets}+2\ell +E_T^{miss}$), where $\ell$ can be either $e$ and/or $\mu$. This final state is illustrated in the Feynman diagram of figure \ref{fig:feynman}.

\subsection{NMSSM-Signal}
\label{sec:signal}


 The allowed NMSSM model parameter spaces are obtained with \texttt{NMSSMTools~5.0.1}  written in SLHA (The Supersymmetry Les Houches Accord)  format, which is then read out by  \texttt{MadGraph v 2.4.3} \cite{Alwall:2014hca} in order to generate the signal events.
The branching ratios
of the Higgs boson and the light top-squark decays, together with all their decay modes, are estimated by using \texttt{NMHDECAY}~\cite{Ellwanger:2004xm} and \texttt{NMSDECAY} \cite{Muhlleitner:2003vg,Das:2011dg},  respectively.

We have generated Monte Carlo samples for integrated luminosities of 300 fb$^{-1}$ for center of mass energies of 13 TeV and 33 TeV.
Signal samples have been generated from leading-order matrix elements using \texttt{MadGraph v 2.4.3} \cite{Alwall:2014hca}. 
The parton level cross-sections at 13 TeV and 33 TeV for the selected benchmark-points are tabulated in Table.\ref{table:nmssmbp}. 

\begin{table}[h!]
\caption{The selected NMSSM benchmark points obtained from \texttt{NMSSMTools~5.0.1}~\cite{Ellwanger:2004xm}
to find the $\tone h_1 \tone$ signal. The values displayed are at the electroweak scale.
The following parameters are fixed: $M_3$ =1900 GeV, $A_{\tau}$ = $A_{\ell}$=1500 GeV and
$M_{\tilde \ell}$ = 300 GeV. Note that we used $M_A$ and $M_P$ as inputs, thus our scenario is not the $Z_3$-NMSSM, and for that $\xi_F$ and $\xi_S$ are non-zero
and also given in the table. We name the product of cross section and branching ratios, $\sigma$ $\times$ BR($h_1 \to b \bar b$) $\times$    
$(\text{BR}(\tone \to b \chiapm)\times\text{BR}(\chiapm \to W^{\pm} \chia))^2$, as $\sigma$.$BR$ for proton collisions at 13(33) TeV.
}
\label{table:nmssmbp}
\begin{center}
\small
\begin{tabular}{l|rrr}
\hline\hline
           & \multicolumn{3}{c}{Benchmark Points}                                  \\ \cline{2-4} 
Parameters & \multicolumn{1}{r}{1} & \multicolumn{1}{r}{2} & \multicolumn{1}{r}{3} 
\\ \hline
$\lambda$&0.328& 0.234 & 0.247
\\
$\kappa$&0.0482& 0.0477&  0.1562
\\
$\tan \beta$&4.053& 9.912& 6.364
\\
$A_\lambda$(GeV)& 1616.55& 1949.17& 2062.75
\\
$A_\kappa$(GeV)& -1011.72& -505.94& 79.71
\\
$\mu_{\rm eff}$& 386.29& 360.90& 387.64
\\
$M_1$ (GeV)& 145.86&131.58& 143.32
\\
$M_2$ (GeV)& 240.22&223.68& 237.69
\\
$M_{\tilde q}$& 671.15& 690.06& 674.24
\\
$A_t$ = $A_b$ (GeV)&-1701.37& -1764.67& -1552.42
\\
$M_A$ (GeV)& 103.44& 117.75& 126.25
\\
$M_P$ (GeV)& 1851.79&1765.12& 1874.95
\\\hline
$\xi_F$ ($10^6$ GeV$^2$)& -1.96&-3.11& -3.61
\\
$\xi_S$ ($10^9$ GeV$^3$)& -3.39&-3.71& -2.06
\\\hline
$m_{\lsp}$ (GeV)& 112.3 & 124.7 & 137.5
\\
$m_{\chiapm}$ (GeV)& 221.3&209.8&222.9
\\
$m_{h_1}$ (GeV)& 70.33 & 62.43 & 64.89
\\
$m_{h_2}$ (GeV)& 125.2& 122.1 & 124.6
\\
$m_{h_3}$ (GeV)& 1683.10&1547.99& 1258.07
\\
$m_{a_1}$ (GeV)&  77.67&  72.23&100.64
\\
$m_{a_2}$ (GeV)& 1850.89& 1764.74& 1874.32
\\
$m_{h^\pm}$ (GeV)& 104.60&106.90& 125.32
\\
$m_{\tone}$ (GeV)&283.8 &293.9& 346.8 
\\ \hline
BR($h_1 \to b \bar b$)&0.911  & 0.912 & 0.910
\\
BR($\tone \to b \chiapm$)& 1.00 &1.00 &1.00
\\
BR($\chiapm \to W \chia$)& 0.834 & 0.960 & 0.863
\\ \hline
$\sigma$ [fb]& 76.98(791.65) & 33.09 (345.71) &17.59(205.95)
\\
$\sigma$.BR[fb]&48.72 (501.03) &27.81(290.57)&11.92(139.58)
\\\hline\hline
\end{tabular}
\end{center}
\end{table}

We have applied some pre-selection requirements at parton level, based on the acceptance and efficiency reconstruction of the LHC detectors,  described as follows.  For jets we require $p_T(j)>15$ GeV and $|\eta(j)|<5$.  For leptons we require $p_T(j)>15$ GeV and $|\eta(j)|<5$. 
The angular distance between jets, leptons, and jets and leptons has been set to $\Delta R > 0.2$, with $\Delta R = \sqrt{\Delta \eta^2 + \Delta \phi^2}$, where $\eta$ and $\phi$ are the pseudo-rapidity and azimuthal angle differences, respectively.
We have used \texttt{Pythia6} \cite{pythia} to generate the parton shower and the hadronization. 

\begin{figure}[ht!]
\centering
\includegraphics[scale=0.52]{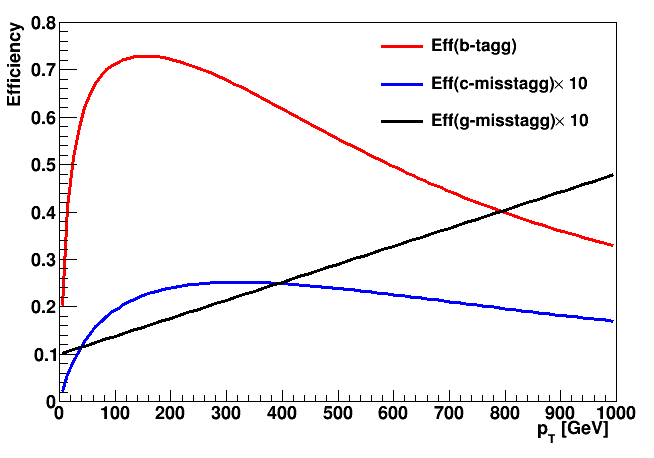}
\caption{B-tagging efficiency as a function of transverse momentum $p_T$.}
\label{fig:btag}
\end{figure}

The detector has been simulated with \texttt{Delphes} \cite{delphes}. For jet reconstruction, the anti-kT algorithm \cite{antikt} in FastJet \cite{fastjet} has been used, with the parameter $R = 0.4$, a $p_T$ threshold of 20 GeV and $|\eta|<$5. 
Electrons (muons) have been isolated using an $R$ parameter of 0.3 (0.4), and $p_T=0.5$ GeV for both muons and electrons.
For the b-tagging efficiency, we have used the default \texttt{Delphes} parameterization for the CMS detector \cite{delphes}. Figure \ref{fig:btag} shows the b-tagging efficiency and gluon and c-jets miss-tagging efficiencies emulated.

\subsection{Backgrounds}
\label{sec:backgrounds}

We consider the SM backgrounds listed in table \ref{tab:backgrounds}, which are in accordance with the CMS and ATLAS studies on SM Higgs production and dileptonic top squark studies \cite{ATLAS:2017tmd, CMS:2016qwm}.
We have generated the Monte Carlo samples at leading-order using \texttt{MadGraph v2.4.3}. For all backgrounds we apply the same parton selection criteria used for the NMSSM-signal events described in the previous subsection. 
We use the NNPDF23LO-PDF set \cite{pdf} for parton distribution functions and the CKKML algorithm \cite{match} for jet matching when required.

We generate the $t\bar{t}$ background with up to two jets at parton level. 
We have divided the $t\bar{t}+2$jets background into two complementary backgrounds which we have named $t\bar{t}+$ two light flavour jets ($t\bar{t}+$l.f.) and $t\bar{t}+$ two heavy flavour jets ($t\bar{t}+$h.f.). Here the light flavour jet is defined as $j=guds$
and the heavy flavour jets contain $b$ and $c$-quarks.
For the case of the $Z+jets$ background, because of its large cross section, we have used a simplified process  of this background where $Z$ is accompanied with a $b\bar{b}$ pair at parton level, matched up to one jet. 
Other backgrounds are top pair production associated with vector bosons $t\bar{t}+V$ with $V=W/Z/\gamma$, top pair production associated with a Higgs $t\bar{t}+H$ and $Wt$.

\begin{table}[h!]
\centering
\caption{SM Backgrounds and cross sections for center of mass energy of 13 and 33 TeV.}
\label{tab:backgrounds}
\begin{tabular}{llrr}
\hline \hline
Background  & Sub-processes   &    $\sigma(13 \text{TeV}) [pb]$    & $\sigma(33 \text{TeV}) [pb]$    \\ 
\hline
$t\bar{t}$+l.f.  & $t\bar{t}+jj$, $(j=guds)$  & 10.02   & 63.62 \\ 
\hline
\multirow{2}{*}{$t\bar{t}$+h.f.} & $t\bar{t}+cj$, $(j=gudsc)$  & 7.072 & 68.07  \\
&$t\bar{t}+bj$, $(j=gudscb)$ & 6.15 & 55.70 \\ 
\hline
\multirow{3}{*}{$t\bar{t}+V$} & $t\bar{t}+W^\pm$  & 0.350  & 1.582  \\
& $t\bar{t}+Z$  & 0.588  & 5.185   \\ 
& $t\bar{t}+\gamma$ & 2.07      & 14.57    \\
\hline
$t\bar{t}+H$    &  & 0.400 & 3.347 \\
\hline
$Wt$    &  & 55.66  & 364.80 \\
\hline
$Z+$jets    & $Z+bbj,(Z\rightarrow \nu\bar{\nu})$ & 13.87 & 64.65  \\ 

\hline \hline
\end{tabular}
\end{table}

\subsection{Event selection }
\label{sec:cuts}

\begin{figure}[ht!]
\centering
\includegraphics[width=\textwidth]{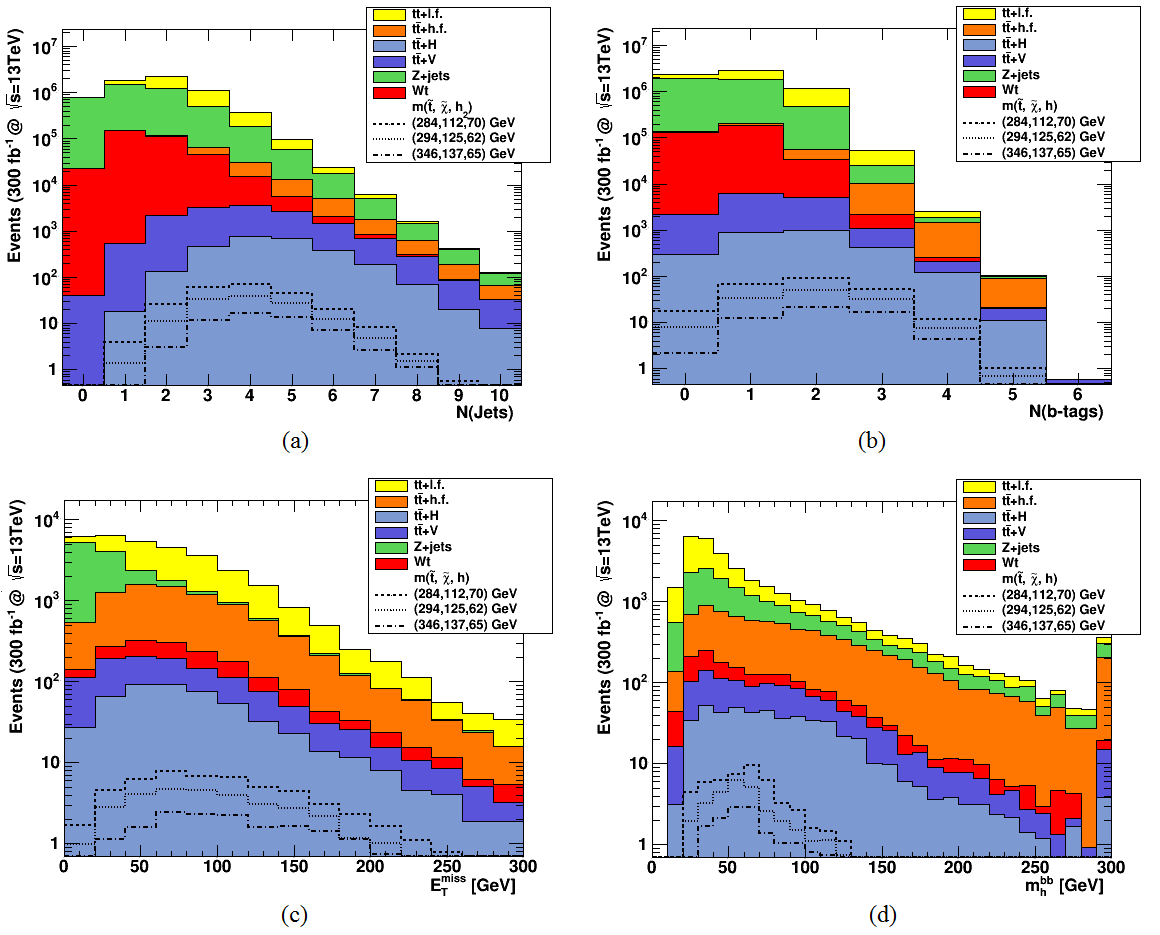}
\caption{Distribution of events at a center of mass energy $\sqrt{s}=$ 13 TeV, for the (a) the number of jets, (b) the number of b-tags, (c) the missing transverse energy and (d) the light Higgs mass reconstruction. Samples are simulated for an integrated luminosity of 300 fb$^{-1}$.  The contributions from all SM backgrounds are shown as histogram stacks and the rightmost bin of each plot includes overflow events.}
\label{fig:plots}
\end{figure}

We require events with exactly two leptons with opposite charge,  with a minimum transverse momentum of 15 GeV for both leptons, and a pseudorapidity of $|\eta(\ell)|<2.4$.
Based on the Feynman diagram of figure \ref{fig:feynman}, we require events with at least four jets. The distributions of events of the number of jets, after the $N(\ell)=2$ requirement, are shown in figures \ref{fig:plots} (a) and \ref{fig:33plots} (a) for 13 TeV and 33 TeV respectively.

After studying the best scenario for signal (S) over background (B)  significance ($S/\sqrt{S+B}$), we  select events with at least three b-tagged jets. The distribution of events of the number of b-tagged jets are shown in figures \ref{fig:plots} (b) and \ref{fig:33plots} (b) for 13 TeV and 33 TeV respectively.
Based on the same significance criterion,
we require a value of missing transverse energy of $E_T^{miss}>$60 GeV. The distribution of events of $E_T^{miss}$ are shown in figures \ref{fig:plots} (c) and \ref{fig:33plots} (c) for 13 TeV and 33 TeV respectively.

Furthermore, we have reconstructed the non-standard Higgs boson invariant mass $m_h^{bb}$ from the two closest b-jets in $\Delta R=\sqrt{\Delta\phi^2+\Delta\eta^2}$. The $m_h^{bb}$ distribution  is shown in figures \ref{fig:plots} (d) and \ref{fig:33plots} (d) for 13 TeV and 33 TeV respectively.
With respect to the reconstructed $m_h^{bb}$ mass , we require a window with an upper value corresponding to the SM Higgs mass of 125 GeV, and a lower value of 40 GeV, obtained from a significance optimization analysis. The distribution of events for the reconstructed mass $m_h^{bb}$ are shown in figures \ref{fig:plots} (d) and \ref{fig:33plots} (d) for 13 TeV and 33 TeV respectively.


After analyzing several kinematic variables suggested in \cite{ATLAS:2017tmd},  we have found that the following ratio variables related to $E_T^{miss}$ are useful to maximize signal significance:

\begin{align}
R_1 =& \frac{E_T^{miss}}
{E_T^{miss}+p_T(\ell_1)+p_T(\ell_2)+
p_T(j_1)+p_T(j_2)}\\
R_{\ell\ell} =& \frac{E_T^{miss}}
{p_T(\ell_1)+p_T(\ell_2)}\\
R_{\ell j} =& \frac{E_T^{miss}}
{E_T^{miss}+p_T(\ell_1)+p_T(\ell_2)+
\sum_{i=1,...,4} p_T(j_i)}
\end{align}

Here $p_T(\ell_1)$ and $p_T(\ell_2)$ are the leading and sub-leading lepton transverse momenta, respectively, and $p_T(j_i)$ are the transverse momenta of jets in decreasing order.
A summary of  the event selection criteria made in our analysis is presented in table \ref{tab:cuts}.

SM backgrounds event flow after each selection requirement is shown in tables \ref{tab:13eventflow} and \ref{tab:33eventflow} for center of mass energies of 13 TeV and 33 TeV, respectively. 
Likewise, event flow for each signal benchmark point is shown in table \ref{tab:eventflowbp}.
Since the significance on the number of signal events ($N=L\sigma$), we also estimate the significance for a higher luminosity of 3000 fb$^{-1}$. This is shown in the bottom row of table \ref{tab:eventflowbp}.

\begin{table}[h!]
\centering
\caption{Event selection criteria for all signal and background samples.}
\label{tab:cuts}
\begin{tabular}{cc}
\hline
\multicolumn{2}{c}{Baseline Cuts}                                                         \\ \hline
\multicolumn{1}{c|}{\multirow{2}{*}{Leptons $(\ell=e,\mu$)}} & $N(\ell)$=2                                  \\
\multicolumn{1}{c|}{}                         & $p_T(\ell)>15$GeV, $|\eta(\ell)|<2.4$ \\ \hline
\multicolumn{1}{c|}{\multirow{3}{*}{Jets}}    & $p_T(j)>20$GeV, $|\eta(j)|<5$   \\
\multicolumn{1}{c|}{}                         & $N(j)\geq$4                        \\
\multicolumn{1}{c|}{}                         & $N(\text{b-tags})\geq$3                    \\ \hline
\multicolumn{1}{l|}{Missing transverse energy}                      & $E_T^{miss}>60$ GeV                           \\ \hline
\multicolumn{2}{c}{Additional cuts}                                                       \\ \hline
\multicolumn{1}{c|}{Mass reconstruction}             & $40<m_h^{bb}<125$ GeV \\
\multicolumn{1}{c|}{$R_{\ell\ell}$}                      & $>0.4$                 \\
\multicolumn{1}{c|}{$R_{\ell j}$}                      &        $>0.1$               \\
\multicolumn{1}{c|}{$R_{1}$}                       &    $>0.14$                                       \\ \hline
\end{tabular}
\end{table}

\begin{table}[ht!]
\centering
\caption{\label{tab:13eventflow} Flow of events at 13 TeV, for each selection  requirement of table \ref{tab:cuts}.}
\small
\begin{tabular}{c|rrrrrr}
\hline\hline 
       & \multicolumn{6}{c}{Backgrounds}                                                                \\ \cline{2-7} 
   & \multicolumn{1}{l}{$t\bar{t}$+l.f.} & \multicolumn{1}{l}{$t\bar{t}$+h.f.}   & \multicolumn{1}{l}{$t\bar{t}+H$} & \multicolumn{1}{l}{$t\bar{t}+V$} & \multicolumn{1}{l}{$Wt$} & \multicolumn{1}{l}{$Z$+jets} \\ \hline
Total Events                 & 6.48$\times 10^{6}$ & 3.96$\times 10^{6}$ & 120 000 & 903 240 & 4.16$\times 10^{6}$ & 1.67$\times 10^{7}$\\
$N(\ell)=2$, ($\ell=e,\mu)$  & 2.85$\times 10^{6}$ & 95239 & 3942 & 26675 & 1.74$\times 10^{6}$ & 432 097\\
$N(j)\geq 4$                 & 227245   & 27867    & 2098 & 8387 & 73108 & 15994\\
$N(\text{b-tags})\geq 3$     & 14813    & 6760     & 511   & 795   & 3093 & 590 \\
$E_T^{miss}>60$ GeV          & 8611     & 4102     & 334   & 435   & 176   & 385 \\
$40<m_{h}^{bb}<125$ GeV      & 3383     & 2185     & 220   & 240   & 103   & 189 \\
$R_{\ell\ell}>0.4$           & 3156     & 2086     & 200   & 210   & 73    & 169 \\
$R_{\ell j}>0.1$             & 3134     & 2040     & 194   & 201   & 54    & 169 \\
$R_{1}>0.14$                 & 3054     & 1990     & 194   & 198   & 45    & 168
\\ \hline \hline
\end{tabular} 
\end{table}

\begin{table}[ht!]
\centering
\caption{\label{tab:33eventflow}Flow of events at 33 TeV, for each selection requirement of table \ref{tab:cuts}.}
\small
\begin{tabular}{c|rrrrrr}
\hline\hline 
       & \multicolumn{6}{c}{Backgrounds}                                                                \\ \cline{2-7} 
   & \multicolumn{1}{l}{$t\bar{t}$+l.f.} & \multicolumn{1}{l}{$t\bar{t}$+h.f.}   & \multicolumn{1}{l}{$t\bar{t}+H$} & \multicolumn{1}{l}{$t\bar{t}+V$} & \multicolumn{1}{l}{$Wt$} & \multicolumn{1}{l}{$Z$+jets} \\ \hline
Total Events                 & 7.50$\times 10^{7}$ & 2.14$\times 10^{7}$ & 1.00$\times 10^{7}$ & 1.95$\times 10^{7}$ & 6.09$\times 10^{7}$ & 3.28$\times 10^{7}$\\
$N(\ell)=2$, ($\ell=e,\mu)$  & 3.29$\times 10^{7}$ & 5.13$\times 10^{5}$ & 33307 & 4.94$\times 10^{5}$ & 2.55$\times 10^{7}$ & 8.49$\times 10^{6}$\\
$N(j)\geq 4$                 & 2.14$\times 10^{6}$   & 1.50$\times 10^{5}$    & 17865 & 1.18$\times 10^{5}$ & 1.07$\times 10^{5}$ & 3.12$\times 10^{5}$\\
$N(\text{b-tags})\geq 3$     & 1.72$\times 10^{5}$    & 36281     & 4311   & 10546   & 45313 & 11793 \\
$E_T^{miss}>60$ GeV          & 99321     & 22022     & 2791   & 6071   & 2583   & 7275 \\
$40<m_{h}^{bb}<125$ GeV      & 38448     & 11726     & 1777   & 3125   & 1514   & 3446 \\
$R_{\ell\ell}>0.4$           & 36035     & 11193     & 1614   & 2769   & 1074    & 3074 \\
$R_{\ell j}>0.1$             & 35602     & 10943     & 1574   & 2672   & 795    & 3074 \\
$R_{1}>0.14$                 & 34767     & 10677     & 1532   & 2614   & 653    & 3052
\\ \hline \hline
\end{tabular} 
\end{table}

\begin{table}[ht!]
\centering
\caption{Event flow for the three signal benchmark points, for each selection requirement of table \ref{tab:cuts}. Significance estimated for 3000 fb$^{-1}$ is shown in the final row. }
\label{tab:eventflowbp}
\begin{tabular}{l|rr|rr|rr}
\hline \hline
 & \multicolumn{6}{c}{$m(\tilde{t}_1,\tilde{\chi}_1^0,h_1^0)$ [GeV]}                                                  \\ \cline{2-7} 
 & \multicolumn{2}{c|}{(284,112,70)} & \multicolumn{2}{c|}{(294,125,62)} & \multicolumn{2}{c}{(346,137,65)} \\ \cline{2-7} 
 & 13 TeV         & 33 TeV        & 13 TeV         & 33 TeV        & 13 TeV        & 33 TeV        \\ \hline
Total Events                    & 14620     & 187600   & 8343 & 87170     & 3576 & 17946 \\
$N(\ell)=2$, ($\ell=e,\mu)$     & 344   & 4427    & 191  & 2019      & 84 & 424 \\
$N(j)\geq 4$                    & 146 & 1870      & 85  & 892      & 42 & 209 \\
$N(\text{b-tags})\geq 3$        & 54    & 691    & 34 & 351       & 19 & 94   \\
$E_T^{miss}>60$ GeV             & 41    &530    & 26    & 269    & 16  & 78  \\
$40<m_{h}^{bb}<125$ GeV    & 30    & 380    & 19 & 199       & 11 & 56   \\
$R_{\ell\ell}>0.4$              & 29    & 372    & 19 & 196       & 11 & 55   \\
$R_{\ell j}>0.1$                & 29    & 369    & 19 & 195      & 11 & 53   \\
$R_{1}>0.14$                    & 29    & 367    & 18 & 192       & 10 & 53 \\
\hline
$S/\sqrt{S+B}$                  & 1.2    & 5.0    & 0.78 & 2.6       & 0.44 & 0.72 
\\ \hline \hline
\end{tabular}
\end{table}

\begin{figure}[ht!]
\centering
\includegraphics[width=\textwidth]{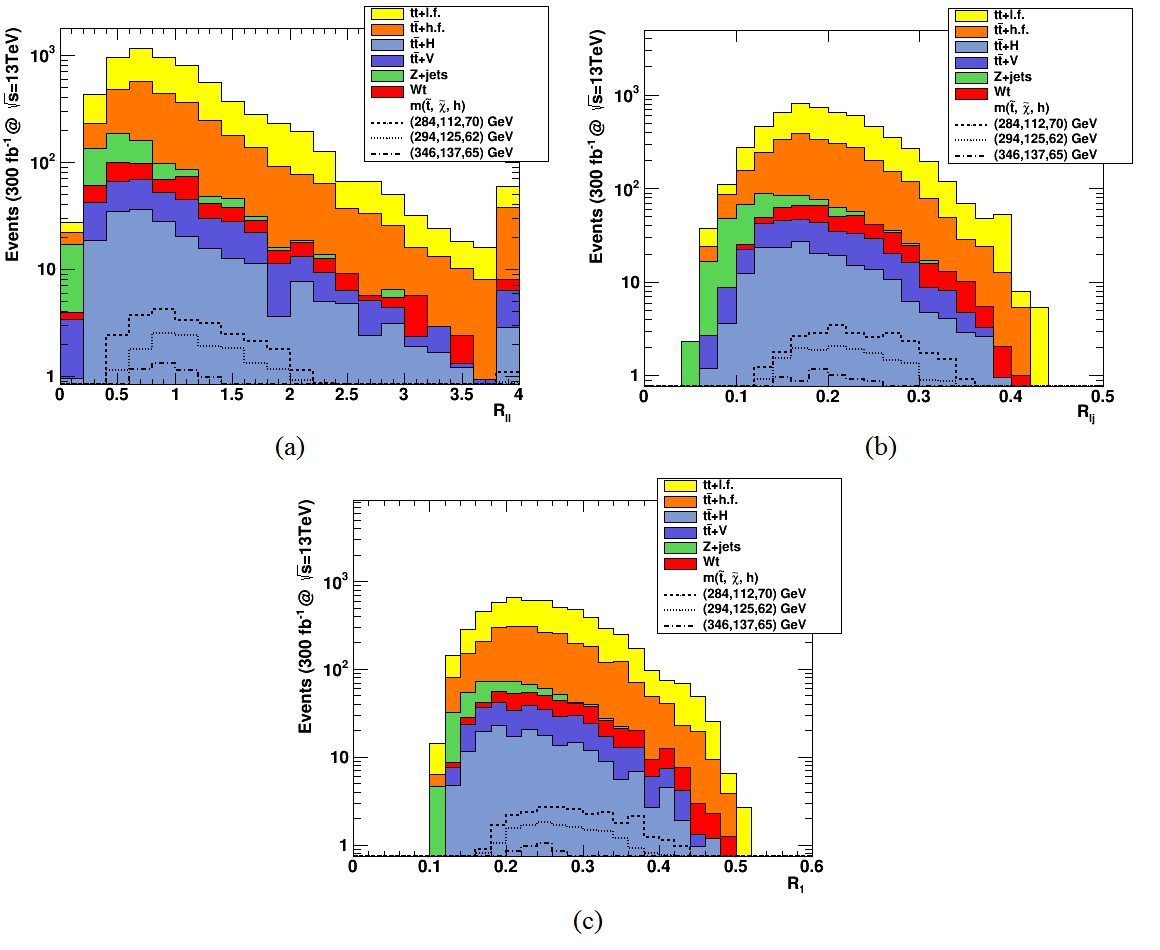}
\caption{Distribution of events at a center of mass energy $\sqrt{s}=$ 13 TeV, for the (a) $R_{\ell\ell}$, (b) $R_{\ell j}$, and (c) the $R_1$ variables. Samples are simulated for an integrated luminosity of 300 fb$^{-1}$.  The contributions from all SM backgrounds are shown as histogram stacks and the rightmost bin of each plot includes overflow events.}
\label{fig:Rs}
\end{figure}

\begin{figure}[ht!]
\centering
\includegraphics[width=\textwidth]{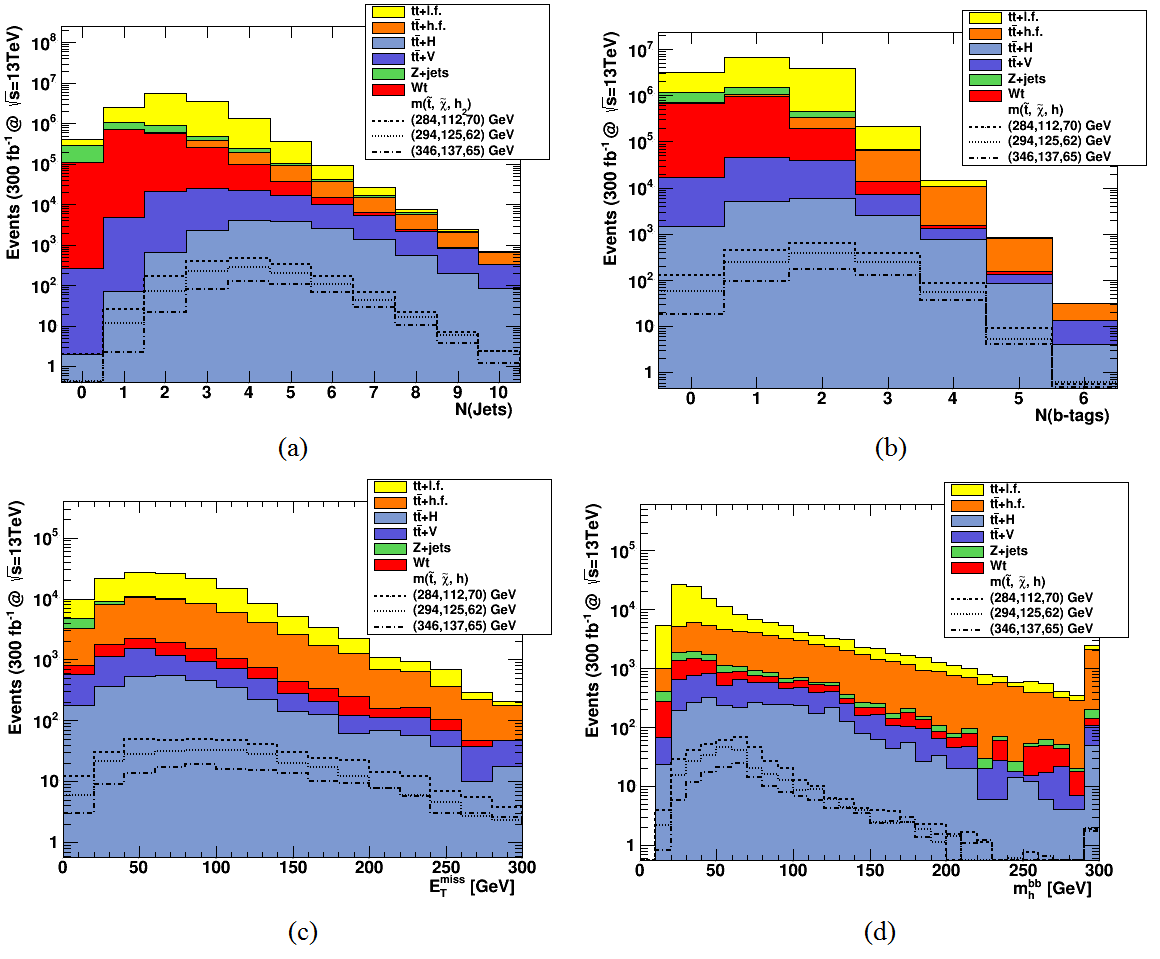}
\caption{Distribution of events at a center of mass energy $\sqrt{s}=$ 33 TeV, for the (a) the number of jets, (b) the number of b-tags, (c) the missing transverse energy and (d) the light Higgs mass reconstruction. Samples are simulated for an integrated luminosity of 300 fb$^{-1}$.  The contributions from all SM backgrounds are shown as histogram stacks and the rightmost bin of each plot includes overflow events.}
\label{fig:33plots}
\end{figure}

\begin{figure}[ht!]
\centering
\includegraphics[width=\textwidth]{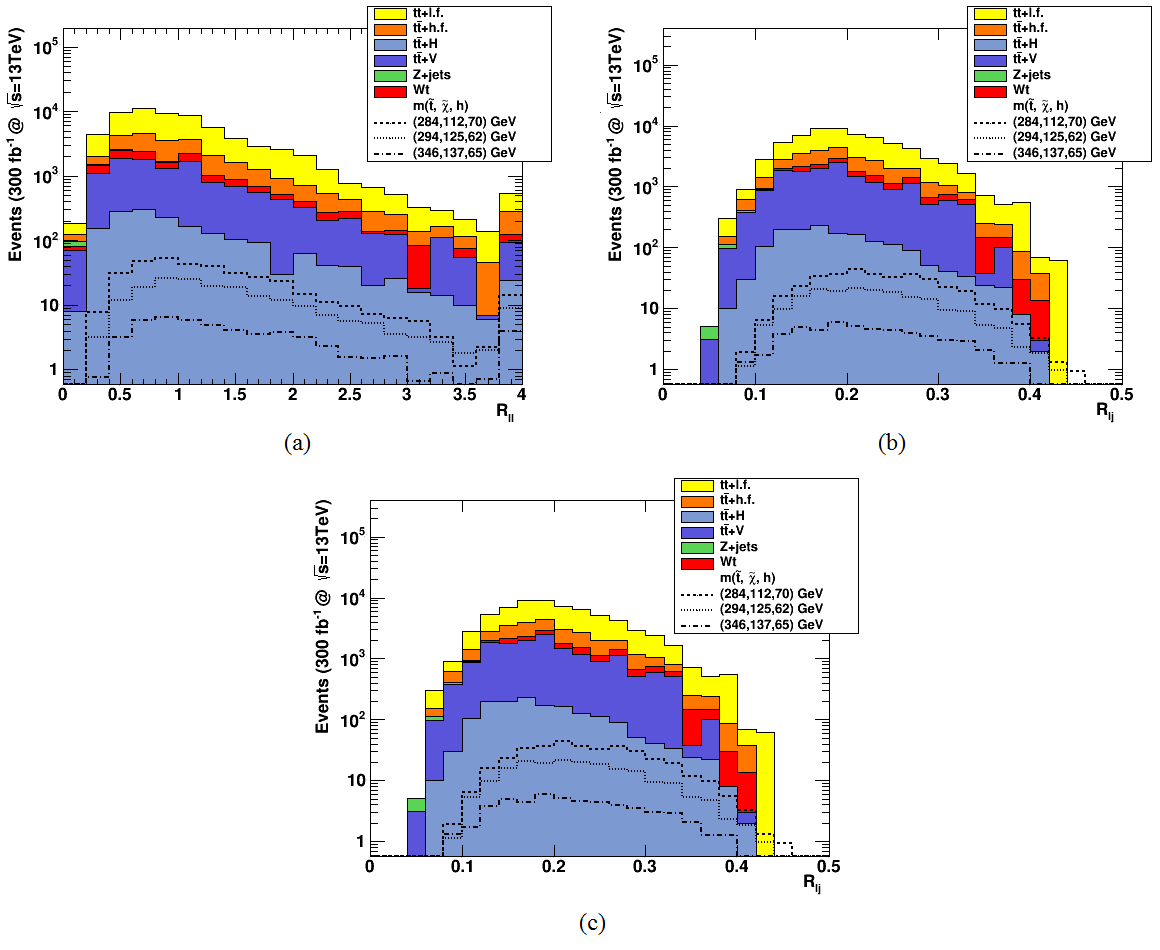}
\caption{Distribution of events at a center of mass energy $\sqrt{s}=$ 33 TeV, for the (a) $R_{\ell\ell}$, (b) $R_{\ell j}$, and (c) the $R_1$ variables. Samples are simulated for an integrated luminosity of 300 fb$^{-1}$.  The contributions from all SM backgrounds are shown as histogram stacks and the rightmost bin of each plot includes overflow events.}
\label{fig:33Rs}
\end{figure}

\subsection{Results}
\label{sec:results}

In this subsection we analyze the performance of different event selections for signal and backgrounds. For the three signal benchmarks we have cross sections  of approximately 0.05-0.01$pb$ (0.5-0.1$pb$), while the total SM backgrounds is approximately 
95$pb$ (640 $pb$), i.e., of the $\order$(3-4) larger than the signal at a proton collide of 13 (33) TeV, respectively. 

After the lepton requirement  we find that approximately 2.4\% of the signal survives, 
 while the backgrounds survive approximately 45\%, 24\%, 
3\%, 2\%, 42\%, 3\% for $t \bar t+$l.f., $t \bar t+$h.f., $t \bar t+$H, $t \bar t+$V, $Wt$ and 
$Z+$jets respectively.
In the next selection, we apply the number of jet requirements ($N_j \geq 4$) and it reduces all backgrounds in large extent and only 7\% pass this criterion. Meanwhile, for the signal, more than 42\% of events are kept. In the next level of 
selections, we apply the $b$-tagged jet criterion ($N(\text{b-tags}) \geq 3$), leaving approximately 7\% of the total background events, while about 40\% of signal events survive. As the signal channels contain at least four hard b-quarks in the parton level, thus higher efficiencies would be preferable.

After we apply the missing transverse energy selections, approximately 53\% of total backgrounds survive, while more than 76\% of signal events are left for benchmark point. 

For the next-level of selection we reconstruct the non-
standard Higgs boson mass from the two $b$-tagged jets with minimum angular distance and the invariant mass ranging between 40-125 GeV. The lower value comes from the distributions, while the upper value is chosen so that the non-standard Higgs boson mass would be lower than the SM-Higgs boson. 
This criterion suppresses the total backgrounds approximately 55\%, while  the signal is reduced by about 27\% . At this, level the number of signal events are around 30-10 (for benchmark points 1 through 3, respectively), while the total backgrounds is 6320. In order to achieve higher significance, we exploit the signal specific selection following ATLAS and CMS studies, namely the ratio variables, $R_{ll}$, $R_{lj}$ and $R_{1}$, as explained in the previous sub-section. 
The $R_{ll}$ selections keep 93\% for total background while for the signal 97\%-100\% 
as expected. The $R_{lj}$  selections keep  the backgrounds approximately 98\%, while no event is reduced for all the three benchmarks points of the signal. 
Finally, the $R_{1}$ selection reduced the total backgrounds approximately 3\%, and for signal it keeps again 100\% of events, except the third benchmark. At the end, we find that we can have a significance of 0.38 $\sigma$ for the benchmark point with masses of the top-squark of 284 GeV and Higgs of 70 GeV. We have estimated the significance for an integrated luminosity of 3000 fb$^{-1}$, obtaining 1.2 $\sigma$ for the same signal benchmark point.

The performance of the individual selections for a higher energy of 33 TeV are somewhat similar  
to the above discussion of 13 TeV. However, as the production cross-sections is larger
at 33 TeV, for the most optimistic benchmark we can have the significance of 5.0$\sigma$ with a luminosity of 3000 fb$^{-1}$.

{
It is interesting to note that the benchmark points 
for which we found the maximal significances are somewhat realized in the so-called compresses scenario, 
where the masses of the lighter top-squark and and mass of the lightest neutralino is 
very close to the top-quark mass. If the LSP is dominantly higgsino type and 
if the mass difference with the lighter top-squark (as Next-to-next-to-LSP) is less than the top-quark 
mass, then the $\tone \to t \lsp$ is kinematically closed and if $\tilde{\chi}_1^{\pm}$   
is NLSP, then naturally the $\tone \to b \chiapm$ decay mode dominates.  And if the mass differences 
between NLSP and LSP is more than the $W$-boson mass, then the whole decay cascades that we consider 
leads to maximal branching ratio factors. This leads to maximal signal rate and hence maximal significances. 
We would also like to add here that in this particular masses scenario, the LSPs co-annihilation to  
the top-quark pair is enhanced by the t-channel lighter top-squark diagram.
}

\section{Summary and Conclusions}

The SM has been very well established after the discovery of the Higgs boson at the LHC in 2012. Adding more than one scalar doublet in a supersymmetric way, without enlarging the gauge 
group, solves some of the SM shortcomings. However, enlarging the scalar doublets leads  to have more scalar bosons with very different mass ranges. But, even enlarging 
the scalar doublet in a supersymmteric way cannot completely explain the requirement 
that the bi-linear Higgs mass-parameter has to be of the order of the electroweak scale. This is called the $\mu$-problem. 
One solution for this problem is to introduce a scalar singlet, and getting VEV of this singlet scalar field, naturally leads to the bi-linear Higgs mass parameters. 
One extra parameter would be adjusted such that it leads to the correct electroweak scale.  
The extra doublet, corresponding to the singlet in the MSSM, is known as the Next-to-MSSM. It enlarges the scalar sectors with different masses and couplings to the SM particles.
Some of the scalar masses would even be lower than the SM-Higgs boson mass of 125 GeV, and are
generally called non-Standard Model Higgs bosons. 

The supersymmetryzation also doubles the particle spectra. The soft-breaking of 
SUSY together with the running of large Yukawa couplings for the 
third generation quarks, makes the lightest top-squark to have a small mass among all the colored sparticles. However, despite the intense search of the lightest top-squark 
at the LHC experiments, there is no signature of its direct evidence. 
The experiments can only quote some lower exclusion limit which is around 95 GeV. 
The difficulties of the non-observation mainly come from its many decay modes, and even with small changes of masses the branching ratios could change abruptly, depending on the model under consideration. 

Hence, looking for any kind of such non-standard Higgses together with the lighter top-squark 
would be a challenge for the present operating LHC and its near future upgrade. 
We are considering this possibility from the associated production of the lighter top-squark 
together with a non-standard Higgs boson, $\tone \tone \hone$ within the NMSSM. The model has naturally 
a low mass lighter neutralino $\lsp$, which serves as a possible candidate for cold dark matter. 
The lighter neutralino is also important as this would appear at the end of the decay chain 
of the lighter top-squark. 

In our analysis, we scanned the NMSSM model parameter spaces using {\tt NMSSMTools v.5.0.1}, assuming that the second intermediate-mass Higgs boson is of the SM-type. For this, we have taken into consideration dark matter constraints (including WMAP), dark matter searches,
B-physics, superparticles mass bounds and the recent Higgs searches results at the LHC experiments. 
In the allowed parameter spaces, we identified the decay patterns of the non-standard Higgs boson 
and the lighter top-squarks. 
We found that the $h_1 \to b \bar b$ channel is dominating in most of the allowed spaces. Also, for low masses of the lighter top-squark, the 
$\tone \to b\chiapm$ channel is predominant 
and the produced chargino mostly decays as $\chiapm \to W \chia$. We assumed the $W$ boson decays into a lepton (electron or muon) for all cases. Considering that both the top-squark decay into identical channels and the $h_1$ productions,   then we have $2\ell+\text{4-jets}+E_T^{miss}$ final states. We estimated the production cross-sections at the LHC and folded with the decay cascade  
branching ratios to get the final event rates. From all the allowed solutions, we have selected three benchmark points with the largest cross sections for our numerical simulations for LHC with present operating energy of 13 TeV and a possible future proton collider with 33 TeV. 

We demand events with at least three jets to be $b$-tagged with proper miss-tagging from light-flavor  and gluon jets. 
We consider the reducible and irreducible SM backgrounds (with charge-conjugation wherever appropriate): 
$t \bar t$+l.f., $t \bar t$+h.f., $t \bar t+H$, $t \bar t+V$, $Wt$, $Z+$jets, where l.f. (h.f.) stands for light (heavy)-flavors and $V$ are the SM-gauge bosons.

We performed a  detector level Monte Carlo simulation using {\tt MadGraph/MadEvent}, {\tt PYTHIA} and {\tt Delphes} within the {\tt Root} analysis framework. We exploited {\tt FastJet} for jet reconstruction using anti-$kT$ algorithm and used the {\tt CKKML} for matching when required. 
The {\tt Delphes} parameters were set in accordance to the CMS detector parameters.

In our event analysis, we first applied selections on the basic event characteristics, like number of leptons, number of jets, number of $b$-jets and missing transverse energy. The values of these selections have been chosen by making an optimization of the significance, defined as $S/\sqrt{S+B}$, of the signal events $S$ with respect to the background $B$ events. 

As in our signal the non-standard Higgs is decaying into two $b$-jets,  we estimated the isolation cone among all possible $b$-tagged jets. The combinations with minimum isolation  angle is the right candidate jets. Then we estimated the invariant masses of these two $b$-tagged jets. 
Then, we applied the window selection $40 < m_{bb} < 125$ GeV, where the lower value has been chosen from a significance optimization of signal rich criterion, and the upper value of 125 GeV, comes from the fact that the non-standard Higgs mass must always be lesser than the SM Higgs boson. 
This particular selection reduces all the SM backgrounds approximately 50\% while the 
signal is reduced by 25\%, and the invariant mass ensures the finding of the non-standard Higgs boson.
At this level, the signal events are approximately 10-30 (60-380) while the total backgrounds are approximately 5500 (52000) for center of mass energies of  13(33) TeV. 

As a next step of our selection, we employed some of the signal specific selections used by the ATLAS and CMS collaborations, namely $R_{ll}$, $R_{lj}$ and $R_{1}$, called ratio variables. Again, by optimizing the Signal rich
and Backgrounds poor, we identified the proper values of their respective selection criteria. We found 
this particular combinations of $R_{ll} \geq 0.4$,  $R_{lj} \geq 0.1$ and $R_{1} \geq 0.14$ are the best 
combinations to suppress SM backgrounds as large as possible, while retain as much as Signal events to 
get the maximal significances. After applying these ratio variables, at LHC with 13 TeV and for the integrated luminosity of 3000 $fb^{-1}$, we have approximately 30-10 signal events for the three benchmark points while the total background is approximately 5650.  
The significances for the three benchmark points are in the range of 1.2 - 0.44 respectively. 
For a 33 TeV proton collider  and for the integrated luminosity of 3000 $fb^{-1}$, we have approximately 369-53 signal events 
for the three benchmark points, while the total background is approximately 53295. The significances for the three 
benchmarks are in the range 5.0 - 0.72 respectively. 

Thus, we conclude that at the LHC with 13 TeV for  an integrated luminosity of  3000 $fb^{-1}$, it is  very difficult to observe the production of the non-standard Higgs boson in association with a pair of light top s-quarks, with significances not high enough to make any exclusion or discovery. However, for a 33 TeV proton collider and  with an integrated luminosity of 3000 fb$^{-1}$, the $\tone\tone\hone$ production of non-standard Higgs bosons with masses up to 70 GeV can be probed with significances of 5.0 $\sigma$.

\section{Acknowledgments}
We thank the administrative department of science, technology and innovation of Colombia (COLCIENCIAS), 
the Faculty of Sciences and the Department of Physics of Universidad de los Andes, Colombia, for the financial support provided to 
this research project. We also acknowledge the computing support provided by the High Performance Computing (HPC) facility of Universidad de los Andes.

\end{document}